\documentclass[sigconf]{acmart}

    \usepackage{booktabs} 

\setcopyright{none} 
    \usepackage{algorithm}
    \usepackage[noend]{algorithmic}

\usepackage{graphicx}

\usepackage{url}

\usepackage{wasysym}

\usepackage{amsthm}
\usepackage{paralist}
\usepackage{endnotes}
\usepackage{enumitem}
\usepackage{multirow}
\usepackage{makecell}
\usepackage{xspace}

\usepackage{pifont}

\newcommand{\code}[1]{{\tt{#1}}}

\newcommand{\paragraphb}[1]{\vspace{0.03in} \noindent{\bf #1} }

\newcommand{\red}[1]{\textcolor{red}{}}

\newcommand{\name}{DeepCorr\xspace}

\newcommand{\nametitle}{D\lowercase{eep}C\lowercase{orr}\xspace}

\graphicspath{{./images/}}

\begin{document}

\date{}

\title{\name:  Strong Flow Correlation Attacks on Tor \\Using Deep Learning}

\author{Milad Nasr \qquad Alireza Bahramali \qquad Amir Houmansadr}
\affiliation{%
  \institution{University of Massachusetts Amherst}
}
\email{{milad,abahramali,amir}@cs.umass.edu}

%


\begin{abstract}

Flow correlation is the core technique used in a multitude of deanonymization attacks on Tor. 
Despite the importance of flow correlation attacks on Tor, existing flow correlation techniques are considered to be ineffective and unreliable in linking Tor flows when applied at a large scale, i.e., they impose  high rates of false positive error rates or require impractically long flow observations to be able to  make reliable correlations. 
In this paper, we show that, unfortunately, flow correlation attacks can be conducted on Tor traffic with drastically higher accuracies  than before by leveraging emerging learning mechanisms. 
We particularly design a system, called \name, that outperforms the state-of-the-art  by significant margins in correlating Tor connections. \name leverages an advanced  deep learning architecture to \emph{learn} a flow correlation function tailored to Tor's complex network\textemdash this is in contrast to  previous works' use of generic statistical correlation metrics to correlated Tor flows. 
We show that with moderate learning, \name can correlate Tor connections (and therefore break its anonymity) with accuracies significantly higher than existing algorithms, and using substantially shorter lengths of flow observations.   For instance, by collecting only about 900 packets  of each target Tor flow (roughly 900KB of Tor data), \name provides a flow correlation accuracy of 96\% compared to 4\% by the state-of-the-art system of RAPTOR using the same exact setting. \red{more numbers}

We hope that our work demonstrates the escalating threat of flow correlation attacks on Tor given recent advances in learning algorithms,  calling for the timely deployment of effective  countermeasures by the Tor community.

\end{abstract}

\maketitle


\section{Introduction}

Tor~\cite{tor-design} is the most widely used anonymity system with more than 2 million daily users~\cite{tor-metrics}. It provides anonymity by relaying clients' traffic through cascades of relays, known as onion-circuits, therefore concealing the association between the IP addresses of the communicating parties. Tor's network comprises around 7,000 public relays, carrying terabytes of traffic every day~\cite{tor-metrics}. Tor is used widely not only by dissidents, journalists, whistleblowers, and businesses, but also by ordinary citizens to achieve anonymity and blocking resistance.    

To be usable for everyday Internet activities like web browsing, Tor aims to provide \emph{low-latency} communications. To make this possible, Tor relays refrain from  obfuscating traffic features like packet timings as doing so will slow down the connections.\footnote{Note that some Tor bridges (but not the public relays)  obfuscate  traffic characteristics of the Tor flows between themselves and censored clients by using various Tor pluggable transports~\cite{pluggable-transport}.}
 Consequently, Tor is known to be susceptible to \emph{flow correlation} attacks~\cite{danezis2004traffic,shmatikov2006timing,murdoch2005low} in which an adversary tries to link the egress and ingress segments of a Tor connection by comparing their traffic characteristics, in particular their packet timings and packet sizes. 

This paper studies  flow correlation attacks on Tor. 
Flow correlation is \textbf{the core technique} used in a wide spectrum of the attacks studied against Tor (and similar anonymity systems)~\cite{insidejob,borisov2007denial,raptor,usersrouted-ccs13,edman2009awareness,starov2016measuring}.
For instance,  in the predecessor attack~\cite{wright2002analysis}  an adversary who controls/eavesdrops multiple Tor relays attempts at deanonymizing  Tor connections by applying flow correlation techniques. The Tor project adopted ``guard'' relays to limit such an adversary's chances of placing herself on the two ends of a target Tor connection. 
 Borisov et al.~\cite{borisov2007denial} demonstrated an active denial-of-service attack that increases an adversary's chances of observing the two ends of a target user's Tor connections (who then performs flow correlation).
 Alternatively,  various routing attacks have been presented on Tor~\cite{raptor,usersrouted-ccs13,edman2009awareness,starov2016measuring} that  aim at increasing an adversary's odds of intercepting the flows to be correlated  by   manipulating the routing decisions. 

Despite the critical role of flow correlation in a multitude of Tor attacks, flow correlating Tor connections has long been considered to be inefficient at scale~\cite{PETs17-TagIt,obfs4-obf,tors-biggest-concern}\textemdash but not anymore! 
Even though Tor relays do not actively manipulate packet timings and sizes to resist flow correlation, the Tor network naturally  perturbs Tor packets  by significant amounts, rendering flow correlation  a difficult problem in Tor. 
Specifically, Tor connections experience large network jitters, significantly larger than  normal Internet connections.
Such large perturbations are resulted by congestion on Tor relays, which is due to the imbalance between Tor's capacity and the bandwidth demand from the clients.   
Consequently, existing flow correlation techniques~\cite{nasr2017compressive,raptor,houmansadr2014non,levine2004timing} suffer from high rates of false positives and low accuracies, unless they are
applied on very long  flow observations and/or impractically small  sets of target flows.
For instance, the state-of-the-art flow correlation of RAPTOR~\cite{raptor} achieves good correlation performance in distinguishing a small set of only 50 target connections, and even this requires the collection of 100 MB over 5 minutes of traffic for each of the intercepted flows.  

In this work, we take flow correlation attacks on Tor to reality. We develop tools that are able to correlate Tor flows with accuracies \emph{significantly higher} than the state-of-the-art\textemdash when applied to large anonymity sets and using very short observations of Tor connections. 
We argue that existing flow correlation techniques~\cite{shmatikov2006timing,raptor,houmansadr2014non,nasr2017compressive,chothia2011statistical,levine2004timing} are inefficient in correlating Tor traffic as they make use of  generic statistical correlation algorithms that are not able to capture the dynamic, complex nature of noise in Tor. 
As opposed to using such  general-purpose statistical correlation algorithms,  in this paper we use deep learning to
\emph{learn a correlation function that is tailored to Tor's ecosystem}.
Our flow correlation system, called \name, then uses the learned correlation function to cross-correlate live Tor flows. 
Note that contrary to website fingerprinting attacks~\cite{panchenko2011website,herrmann2009website,cai2012touching,wang2013improved,wang2014effective}, \name does \emph{not} need to learn any target  destinations or   target circuits; instead \name learns a correlation function that can be used to link flows on \emph{arbitrary circuits}, and to \emph{arbitrary destinations}. In other words, \name can correlate the two ends of a Tor connection even if the connection destination has not been part of the learning set. Also, \name can correlate flows even if they are sent over Tor circuits different than the circuits used during the learning process. 
This is possible as \name's neural network learns the \emph{generic} features of  noise in Tor, regardless of the specific circuits and end-hosts used during the training process.  


We demonstrate \name's strong performance through large scale experiments on live Tor network.
We browse the top 50,000 Alexa websites over Tor, and evaluate \name's true positive and false positive rates in correlating the ingress and egress segments of the recorded Tor connections. 
To the best of our knowledge, 
our dataset is the largest dataset of correlated Tor flows, which we have made  available to the public.
Our experiments show that \name can correlate Tor flows with accuracies \emph{significantly superior} to existing flow correlation techniques. 
For instance, compared to the state-of-the-art flow correlation algorithm of RAPTOR~\cite{raptor}, \name offers a correlation accuracy\footnote{To be fair, in our comparison with RAPTOR we derive the accuracy metric similar to RAPTOR's paper~\cite{raptor}: each flow is paired with only one flow out of all evaluated flows. For the rest of our experiments, each flow can be declared as correlated with arbitrary number of intercepted flows, which is a more realistic (and more challenging) assumption.} of 96\% compared to RAPTOR's accuracy of 4\% (when both collect 900 packets of traffic from each of the intercepted flows)! 
\red{add more}
The following is a highlight of \name's performance:
\begin{itemize}
	\item We train \name using 25,000 Tor flows generated by ourselves. Training \name  takes about a day on a single TITAN X GPU, however we show that an adversary needs to re-train \name only about once every month to preserve its  correlation performance.  
\item	\name can be used as a generic correlation function: \name's performance is consistent for various test datasets  with different sizes and containing flows routed over different circuits. 
	\item \name outperforms prior flow correlation algorithms by very large margins. Importantly, \name enables the correlation of Tor flows with flow observations much shorter than what is needed by previous work. For instance, with only 300 packets, \name achieves a true positive rate of 0.8 compared to  less than 0.05 by prior work (for a fixed false positive rate of $10^{-3}$). 
	\item \name's performance rapidly improves with longer flow observations and with larger training sets. 
	\item \name's correlation time is significantly faster than previous work for the same target accuracy. For instance, each \name correlation takes 2ms compared to RAPTOR's more than 20ms, when both target a 95\% \red{accuracy} on identical  dataset. 
\end{itemize}

%

We hope that our study raises concerns in the community on the escalating risks of large-scale traffic analysis on Tor communications in light of the emerging deep learning algorithms. A possible countermeasure to \name is deploying traffic obfuscation techniques, such as those employed by Tor pluggable transports~\cite{pluggable-transport}, on \emph{all} Tor traffic. 
  We evaluate the performance of \name on each  of Tor's currently-deployed pluggable transports, showing that meek and obfs4-iat0 provide little protection against \name's flow correlation, while obfs4-iat1 provides a better protection against \name (note that none of these obfuscation mechanisms are currently deployed by \emph{public} Tor relays, and even  obfs4-iat1 is deployed by a small fraction of Tor bridges~\cite{obfs4-obf}). This calls for designing effective traffic obfuscation mechanisms to be deployed by Tor relays  that do not impose large bandwidth and performance overheads on Tor communications.

Finally, note that while we present \name as a flow correlation attack on Tor, it can be used to correlate flows in other flow correlation applications as well. 
To demonstrate this, we also apply  \name to the problem of 
stepping stone detection~\cite{he2007detecting,wang2003robust,blum2004detection}
showing that  \name significantly outperforms previous stepping stone detection algorithms in unreliable network settings. 

\paragraphb{Organization:} The rest if this paper is organized as follows. \red{dfdfdf}

\section{Preliminaries and Motivation}\label{sec:related}

Flow correlation attacks, also referred to as \emph{confirmation attacks},  are used to \emph{link network flows} in the presence of encryption 
and other content obfuscation mechanisms~\cite{zhang:sec00,wang:esorics02,donoho2002multiscale,he2007detecting,danezis2004traffic,shmatikov2006timing,Ling2009,nasr2017compressive}. 
In particular, flow correlation techniques can break anonymity in anonymous communication systems like  Tor~\cite{tor-design} and mix networks~\cite{morphmix,danezis2003mixminion,reiter1998crowds} by linking the egress and 
ingress segments of the anonymous connections through correlating traffic features~\cite{back:ih2001,zhu:pet05,ramsbrock2008first,wang:oakland07,Wang2005,murdoch2005low,
danezis2004traffic,shmatikov2006timing}.
Alternatively, flow correlation techniques can be used to identify  
cybercriminals who use network proxies to obfuscate their identities, i.e., stepping stone attackers~\cite{staniford-chen:oakland95,yoda:esorics00,zhang:sec00}.

\subsection{Threat Model}
\begin{figure*}[!t]
    \centering
    \includegraphics[scale=0.3]{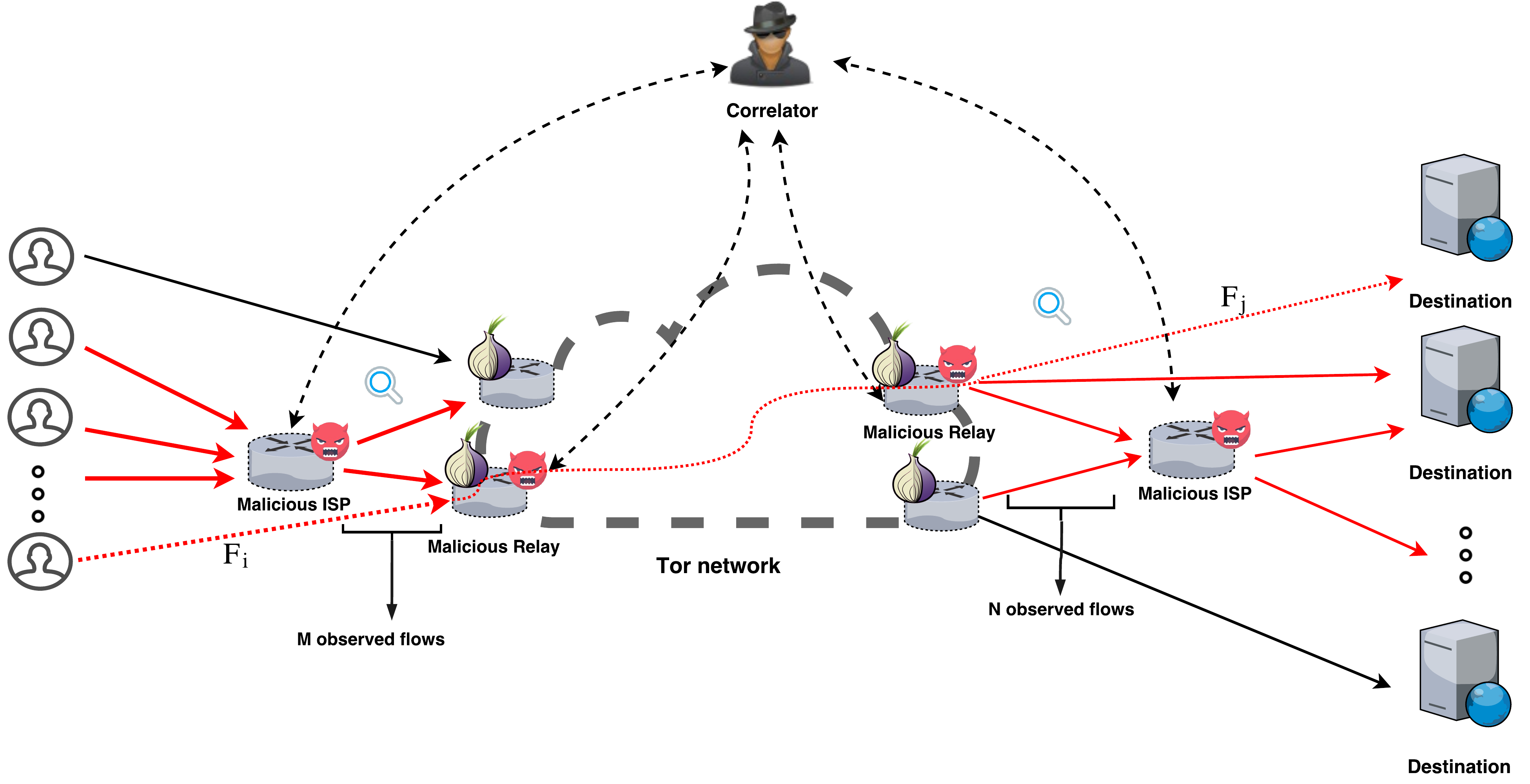}
    \caption{The main setting of a flow correlation attack on Tor. The adversary intercepts Tor flows either by running  malicious Tor relays or wiretapping  Internet ASes and IXPs.}
    \label{fig:threat}
\end{figure*}

Figure~\ref{fig:threat} shows the main setting of a flow correlation scenario. 
The setting consists of a computer network (e.g., Tor's network)  with $M$ ingress flows and $N$ egress flows. Some of the egress flows are the obfuscated versions of some of the ingress flows; however, the relation between such flows  can not  detected using packet contents   due to the use of encryption and similar content obfuscation techniques like onion encryption. For instance, in the case of Tor, $F_i$ and $F_j$ are the entry and exit segments of one  Tor connection (see Figure~\ref{fig:threat}), however, such association can not be detected by inspecting the packet contents of   $F_i$ and $F_j$ due to onion encryption. We call $(F_i,F_j)$ a pair of \emph{associated flows}.

The goal of an adversary in this setting is to identify (some or all of) the associated flow pairs, e.g., $(F_i,F_j)$, by comparing  traffic characteristics, e.g., packet timings and sizes, across all of the ingress and egress flows. 
Linking associated flow pairs using traffic characteristics is called flow correlation.

A flow correlation adversary can intercept network flows  at various network locations. A Tor adversary, in particular, can intercept Tor flows
either by running  malicious Tor relays~\cite{borisov2007denial,insidejob,wright2002analysis}  or by controlling/wiretapping  Internet ASes or IXPs~\cite{johnson2013users,raptor,starov2016measuring}. We further elaborate on this in Section~\ref{sec:Tor-attacks}.

Note that in this paper we study \emph{passive} flow correlation attacks only; therefore, \emph{active} flow correlation techniques, also known as flow watermarks as    introduced in Section~\ref{sec:other}, are out of the scope of this paper. Also, flow correlation is different from website fingerprinting attacks, as discussed in Section~\ref{sec:other}.





\subsection{Existing Flow Correlation Techniques}\label{sec:prev-tech}


As mentioned before, flow correlation techniques use traffic features, particularly, packet timings, packet sizes, and their variants (e.g., flow rates, inter-packet delays, etc.), to correlate and link network flows (recall that packet contents can \emph{not} be used to link flows in this setting due to content obfuscation, e.g.,  onion encryption). 
For instance, the early work of Paxson and Zhang~\cite{zhang:sec00} models packet arrivals as a series of ON and OFF patterns, 
which they use to correlate network flows, and Blum et al.~\cite{blum:raid04} correlate the aggregate sizes of network packets over time. 
Existing flow correlation techniques mainly use \emph{standard statistical correlation metrics} to correlate the vectors of flow timings and sizes across flows. 
In the following, we overview the major types of statistical correlation metrics used by previous flow correlation algorithms.

%

\paragraphb{Mutual Information}
The mutual information metric  measures the dependency of two random variables. It, therefore,  can be used to quantify the correlation of  flow features across flows, e.g., the traffic features of an egress Tor flow depends on the features of its corresponding ingress flow. The mutual information technique has been used by Chothia et al.~\cite{chothia2011statistical} and Zhu et al.~\cite{zhu2004flow} to link flows. This metric, however, requires a long vector of features (e.g., long flows) in order to make reliable decisions, as it needs to reconstruct and compare the empirical distributions of traffic features of  target flows.

\paragraphb{Pearson Correlation}
The Pearson Correlation coefficient is a classic statistical metric for linear correlation between random variables. Unlike the mutual information metric, the Pearson Correlation metric does not need to build the empirical distribution of the variables it is correlating, and therefore can be applied on a shorter length of data. The Pearson Correlation metric has been used by several flow correlation systems~\cite{levine2004timing,shmatikov2006timing}.

\paragraphb{Cosine Similarity}
The Cosine similarity metric measures the angular similarity of two random variables. Similar to the Pearson coefficient, it can be directly applied on  the sample vectors  of two random variables. 
This metric has been used by different timing and size correlation systems~\cite{nasr2017compressive,houmansadr2014non} to link network flows. 

\red{add more houmansa citations}

\paragraphb{Spearman Correlation}
The Spearman rank correlation metric measures the statistical dependence between the rankings of two variables. The metric can be defined as the Pearson correlation between  ranked variables.
The recent work of RAPTOR~\cite{raptor} uses this metric to correlate Tor flows.



\subsection{Flow Correlation Attacks on Tor}\label{sec:Tor-attacks}

Flow correlation is the \emph{core technique} used in a broad range of attacks studied against Tor (and other anonymity systems). 
To be able to perform flow correlation, an adversary needs to observe (i.e., intercept) some fraction of flows entering and exiting the Tor network. The adversary can then  deanonymize a specific Tor connection, if she is able to intercept both of the ingress and egress segments of that Tor connection (by performing a flow correlation algorithm on those flow segments). 
Therefore, an adversary can increase her chances of deanonymizing Tor connections by trying to intercept a larger fraction of Tor's  ingress and egress flows.

There are two main approaches  an attacker can take to  increase the fraction of Tor connections she is intercepting.
First, by running a large number of Tor relays and recording the traffic features of the Tor connections they relay. 
Various studies have shown that an adversary with access to such malicious relays can increase her chances of intercepting the both ends of a Tor connection in different ways~\cite{arnbak2014loopholes,wright2002analysis,borisov2007denial,mittal2011stealthy,hopper2010much}. For instance, Borisov et al.~\cite{borisov2007denial} demonstrate an active denial-service-attack to increase the chances of intercepting the ingress and egress segments of a target client's Tor traffic. The Tor project has adopted the concept of Tor guard relays~\cite{changeGuards} to reduce the chances of performing flow correlation by an adversary controlling malicious relays, an attack known as the predecessor attack~\cite{wright2002analysis}.

Alternatively, an adversary can increase her opportunities of performing flow correlation
by controlling/wiretapping  autonomous systems (ASes) or Internet exchange points (IXPs), and recording the traffic features of the  Tor connections that they transit. 
 Several studies~\cite{feamster:wpes04,raptor,murdoch:pet07} demonstrate that specific ASes and IXPs intercept a significant fraction of Tor traffic, therefore are capable of performing flow correlation on Tor at large scale.  
Others~\cite{raptor,usersrouted-ccs13,starov2016measuring,johnson2013users,edman2009awareness} show that an AS-level adversary can further increase her chances of flow correlation by performing various routing manipulations that reroute a larger fraction of Tor connections through her adversarial ASes and IXPs.  
For instance, Starov et al.~\cite{starov2016measuring} recently  show  that approximately 40\% of Tor circuits are vulnerable to flow correlation attacks by a single malicious AS, and Sun et al.~\cite{raptor} show that churn in BGP as well as active manipulation of  BGP updates can amplify an adversarial AS's visibility on Tor connections.
This has lead to various proposals on deploying AS-aware path selection mechanisms for Tor~\cite{edman2009awareness,akhoondi2012lastor,nithyanand2015measuring}.




\subsection{This Paper's Contributions}

While flow correlation is the core of a multitude of  attacks on Tor~\cite{raptor,usersrouted-ccs13,starov2016measuring,johnson2013users,edman2009awareness,arnbak2014loopholes,wright2002analysis,borisov2007denial,mittal2011stealthy,hopper2010much,feamster:wpes04,raptor,murdoch:pet07,nithyanand2015measuring}, existing flow correlation algorithms are assumed to be ineffective in linking Tor connections reliably and at scale~\cite{PETs17-TagIt,obfs4-obf,tors-biggest-concern}. This is due to Tor's extremely noisy  network that applies large perturbations on  Tor flows, therefore rendering  traffic features across associated ingress and egress Tor flows hard to get reliably correlated. 
In particular, Tor's network applies large network jitters on Tor flows, which is due to congestion on Tor relays, and many Tor packets are fragmented and repacketized due to unreliable network conditions. 
Consequently, existing flow correlation techniques offer poor correlation performances\textemdash unless applied to very large flow observations as well as  unrealistically small sets of target flows.\footnote{Note that  \emph{active} attacks like \cite{shmatikov2006timing} are out of our scope, as discussed in Section~\ref{sec:other}, since such attacks are  easily detectable, and therefore can \emph{not} be deployed by an adversary at large scale for a long time period without being detected.} 
For instance, the  state-of-the-art correlation technique of Sun et al.~\cite{raptor} needs to observe 100MB of traffic from each target flow for around 5 minutes to be able to perform  reliable flow correlations. Such long flow observations  not only are impractical due to the  short-lived nature of typical Tor connections (e.g., web browsing sessions), but also impose unbearable storage requirements if applied at large scale (e.g., a malicious Tor relay will likely intercepte tens of thousands of concurrent flows). 
Moreover, existing techniques suffer from high rates of false positive correlations unless applied on an unrealistically small set of suspected flows, e.g.,  Sun et al.~\cite{raptor} correlate among a set of only 50 target flows. 


\paragraphb{Our Approach:}
We believe that the main reason for the ineffectiveness of existing flow correlation techniques is the intensity as well as the unpredictability of network perturbations in Tor. 
We argue that previous flow correlation techniques are inefficient in correlating Tor traffic since they make use of  general-purpose statistical correlation algorithms that are not able to capture the dynamic, complex nature of noise in Tor. 
As opposed to using such  generic statistical correlation metrics, in this paper we use deep learning to
\emph{learn a correlation function that is tailored to Tor's ecosystem}.
We design a flow correlation system, called \name, that learns a flow correlation function for Tor, and uses the learned function to cross-correlate live Tor connections. 
Note that contrary to website fingerprinting attacks~\cite{panchenko2011website,herrmann2009website,cai2012touching,wang2013improved,wang2014effective}, \name does \emph{not} need to learn any target  destinations or target circuits; instead \name learns a correlation function that can be used to link flows on \emph{arbitrary circuits}, and to \emph{arbitrary destinations}. In other words, \name can correlate the two ends of a Tor connection even if the connection destination has  not been part of the learning set. Also, \name can correlate flows even if they are sent over Tor circuits different than the circuits used during the training process.  

We demonstrate \name's strong correlation performance through large scale experiments on live Tor network, which we compare to previous flow correlation techniques.
We hope that our study raises concerns in the community on the increasing risks of large-scale traffic analysis on Tor in light of emerging  learning algorithms. 
We discuss  potential countermeasures, and evaluate \name's performance against existing countermeasures.

\subsection{Related Topics Out of Our Scope}\label{sec:other}

\paragraphb{Active flow correlation (watermarking)}
Network flow watermarking is an \emph{active} variant of the flow correlation techniques introduced above.
Similar to passive flow correlation schemes, flow watermarking  aims at linking network flows using traffic features that persist  content obfuscation, i.e., packet sizes and timings. 
By contrast, flow watermarking systems need to \emph{manipulate} 
the traffic features of the flows they intercept in order to be able to perform flow correlation. 
In particular, many flow watermarking systems~ \cite{houmansadr:ndss09,pyun:infocom07,icassp11,wang:oakland07,yu:oakland07,houmansadr2013need,swirl}
perturb packet timings of the intercepted flows by slightly delaying network 
packets  to modulate an artificial pattern into the flows, called the watermark. 
For instance, RAINBOW~\cite{houmansadr:ndss09} manipulates the 
inter-packet delays of network packets in order to embed a watermark signal. 
Several proposals~\cite{pyun:infocom07,houmansadr:icassp09,wang:oakland07,yu:oakland07,kiyavash:sec08}, 
known as interval-based watermarks, 
work by delaying packets into  secret time intervals. 

While passive flow correlation attacks (studied in this paper) are information theoretically undetectable, a watermarking adversary may reveal herself by applying traffic perturbations that differ from that of normal traffic. Some active correlation techniques~\cite{chakravarty2014effectiveness,shmatikov2006timing} do not even aim for invisibility, therefore they can be trivially detected and disabled, making them unsuitable for large scale  flow correlation. 
Additionally, while passive flow correlation algorithms can be computed offline, flow watermarks need to  be performed by resourceful adversaries who are able to apply traffic manipulations on live Tor connections.  
In this paper, we only focus on passive flow correlation techniques.

\paragraphb{Website Fingerprinting}
Website fingerprinting attacks~\cite{panchenko2011website,herrmann2009website,cai2012touching,wang2013improved,wang2014effective,juarez2014critical,lu2010website,panchenko2016website,hayes2016k,wang2016realistically,he2014novel} use a different threat model than flow correlation techniques. In website fingerprinting, an adversary intercepts a target client's ingress Tor traffic (e.g., by wiretapping the link between a Tor client and her guard relay), and compares the intercepted ingress Tor connection to the  traffic fingerprints of a finite (usually small) set of target websites. This is unlike flow correlation attacks in which the adversary intercepts the \emph{two ends} of an anonymous connection, enabling the attacker to deanonymize  \emph{arbitrary} senders and receivers. Existing website fingerprinting systems leverage standard machine learning algorithms such as SVM and kNN to classify and identify target websites, and recent work~\cite{DL-WF-18} has investigated the use of deep learning for website fingerprinting. In contrary, as overviewed in Section~\ref{sec:prev-tech}, prior passive flow correlation techniques use  statistical correlation metrics to link traffic characteristics across network flows. 
We consider website fingerprinting orthogonal to our work as it is based on  different threat model and techniques. 
 

%
%
%
%
%
%
%
%
%
%

\section{Introducing \nametitle}

In this section, we introduce our flow correlation system, called \name, which uses deep learning algorithms to learn correlation functions. 



\subsection{Features and Their Representation}

Similar to existing flow correlation techniques overviewed earlier, our flow correlation system uses the timings and sizes of network flows to cross-correlate them. 
A main advantage~\cite{goodfellow2016deep} of deep learning algorithms over conventional learning techniques is that a deep learning model can be provided with \emph{raw} data features as opposed to engineered traffic features (like those used by SVM- and kNN-based website fingerprinting techniques~\cite{panchenko2011website,herrmann2009website,cai2012touching,wang2013improved,wang2014effective,lu2010website,panchenko2016website,hayes2016k,he2014novel}).
This is because deep learning is able to  extract complex, effective features  from the raw input features~\cite{goodfellow2016deep} itself. 
Therefore, \name takes raw flow features as input, and uses them to derive complex features, which is used by its correlation function.

We represent a bidirectional network flow, $i$, with the following array:
\[
F_i=[T^u_{i};S^u_{i};T^d_{i};S^d_{i}]
\]
where $T$ is the vector of inter-packet delays (IPD) of the flow $i$, $S$ is the vector of $i$'th packet sizes, and the $u$  and $d$ superscripts represent ``upstream'' and ``downstream'' sides of the bidirectional flow $i$ (e.g., $T^u_{i}$ is the vector of upstream IPDs of $i$). Also, note that we only use the first $\ell$ elements of each of the  vectors, e.g., only the first $\ell$ upstream IPDs. If a vector has fewer than $\ell$ elements, we pad it to $\ell$ by appending zeros. We will use the flow representation $F_i$ during our learning process.

Now suppose that we aim at correlating two flows $i$ and $j$ (say $i$ was intercepted by a malicious Tor guard  relay and $j$ was intercepted by an accomplice exit relay). 
We represent this pair of flows with the following two-dimensional array composed of 8 rows:
\[
F_{i,j}=[T^u_{i}; T^u_{j}; T^d_{i}; T^d_{j}; S^u_{i}; S^u_{j}; S^d_{i}; S^d_{j}]
\] 
where the lines of the array are taken from the flow representations $F_i$ and $F_j$.



\subsection{Network Architecture}

\begin{figure*}[!t]\centering
    \includegraphics[scale=0.16]{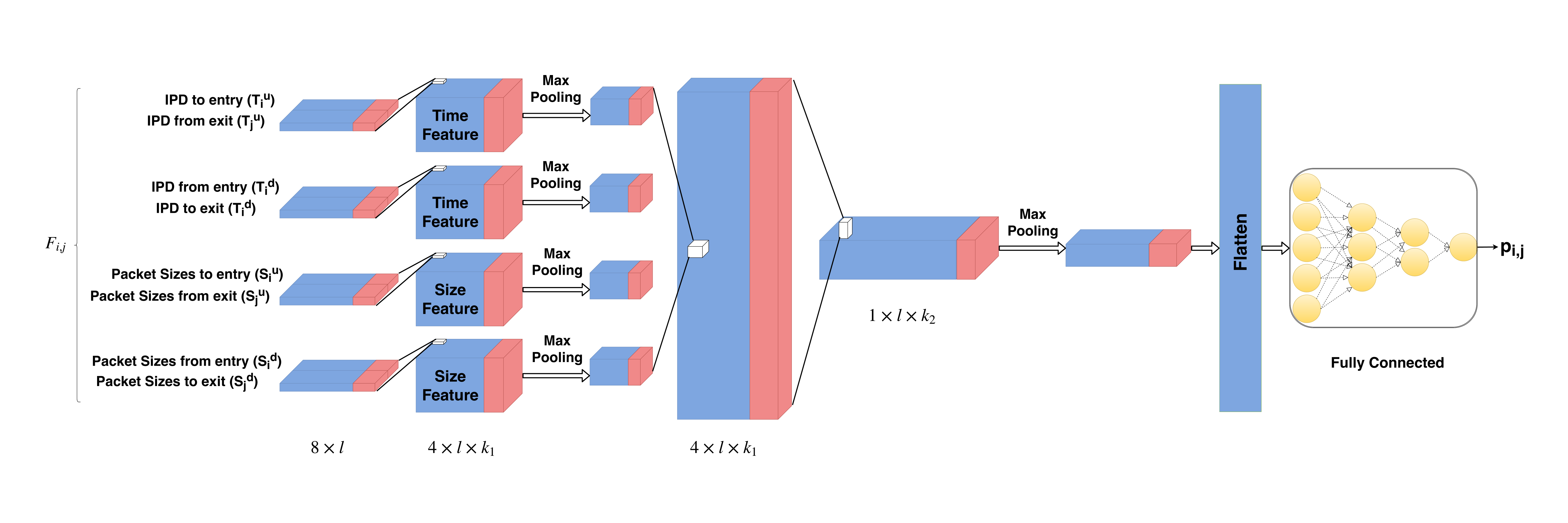}
    \caption{The network architecture of \name. }
    \label{fig:nn_design}
\end{figure*}

We use a Convolutional Neural Network (CNN)~\cite{goodfellow2016deep} to learn a correlation function for Tor's noisy network. 
We use a CNN
 since network flow features can be modeled as time series,  and the CNNs are known to have good performance on time series~\cite{goodfellow2016deep}.
Also, 
the CNNs are invariant to the position of the patterns in the data stream~\cite{goodfellow2016deep}, which makes them ideal to look for possibly shifted traffic patterns.\footnote{Note that our work is the \emph{first} to use a learning mechanism for flow correlation. In our search of effective learning mechanisms for flow correlation, we tried various algorithms including fully connected neural networks, recurrent neural network (RNN), and support vector machine (SVM). However, CNN provided the best flow correlation performance compared to all the other algorithms we investigated, which is intuitively because CNNs are known to work better for longer data lengths. For instance, we  achieved an accuracy of only $0.4$ using fulling-connected  neural networks, which is significantly lower than our performance with CNNs.  
}  

Figure~\ref{fig:nn_design} shows the structure of \name's CNN network. 
The network takes a flow pair $F_{i,j}$ as the input (on the left side). 
\name's architecture is composed of two layers of convolution and three layers of a fully connected neural network. The first convolution layer has $k_1$ kernels each of  size $(2,w_1)$, where $k_1$ and $w_1$ are the hyperparameters, and we use a stride of $(2,1)$.
The intuition behind using the first convolution layer is to capture correlation between the adjacent rows of the input matrix $F_{i,j}$, which  are supposed to be correlated for associated Tor flows, e.g., between $T_i^u$ and $T_j^u$. 

\name's second convolution layer aims  at capturing traffic features from the combination of all timing and size features. At this layer,  \name uses $k_2$ kernels each of size $(4,w_2)$, where $k_2$ and $w_2$ are also our hyperparameters, and it uses a stride of $(4,1)$.
 
The output of the second convolution layer is  flattened and fed to a fully connected network with three layers.
\name uses max pooling after each layer of convolution to ensure permutation invariance and to avoid overfitting~\cite{goodfellow2016deep}.
 Finally, the output of the network is:
  \[
p_{i,j}=  \Psi(F_{i,j})
  \]
   which is used to decide if the two input flows in $F_{i,j}$ are  correlated or not. 
   To normalize the output of the network, we apply a sigmoid function~\cite{goodfellow2016deep} that scales the output between zero and one. 
   Therefore, $p_{i,j}$ shows the  probability of the flows $i$ and $j$ being associated (correlated), e.g., being the entry and exit segments of the same Tor connection. 
   
   \name declares the flows $i$ and $j$ to be correlated if $p_{i,j}>\eta$, where $\eta$ is our \emph{detection threshold} discussed during the experiments. 
   
  The  parameters $(w_1,w_2,k_1,k_2)$ are the hyperparameters of our system; we will tune their values through experiments.

\subsection{Training}

To train our network, we use a large set of  flow pairs that we created over Tor. 
This includes a large set of associated  flow pairs, and a large set of non-associated  flow pairs. 
An associated  flow pair, $F_{i,j}$, consists of the two segments of a Tor connection (e.g., $i$ and $j$ are the ingress and egress segments of a Tor connection). We label an associated  pair with $y_{i,j}=1$. 
On the other hand,  each non-associated   flow pair (i.e., a negative sample) consists of two arbitrary Tor flows that do not belong to the same Tor connection. We label such non-associated pairs with $y_{i,j}=0$. 
For each captured Tor entry flow, $i$, we create $N_{neg}$  negative samples by forming $F_{i,j}$ pairs where $j$ is  the exit segment of an arbitrary Tor connection. 
$N_{neg}$ is a hyperparameter whose value will be obtained through experiments. 

Finally, we define \name's loss function  using a cross-entropy function as follows:
\begin{equation}
\mathcal{L}=- \frac{1}{|\mathcal{F}|} \sum_{F_{i,j}\in \mathcal{F}} y_{i,j} \log \Psi(F_{i,j}) + (1-y_{i,j}) \log (1-\Psi(F_{i,j}))
\label{eq:loss}    
\end{equation}
 where $\mathcal{F}$ is our training  dataset, composed of all  associated and non-associated flow pairs. We used the Adam optimizer~\cite{kingma2014adam} to minimize the loss function in our experiments. 
The learning rate of the Adam optimizer is another  hyperparameter of our system.

\section{Experimental Setup}
In this section, we discuss our data collection and its ethics, the choice of our hyperparameters, and our evaluation metrics.
 

\subsection{Datasets and Collection}

\begin{figure*}
    \centering
    \includegraphics[scale=0.6]{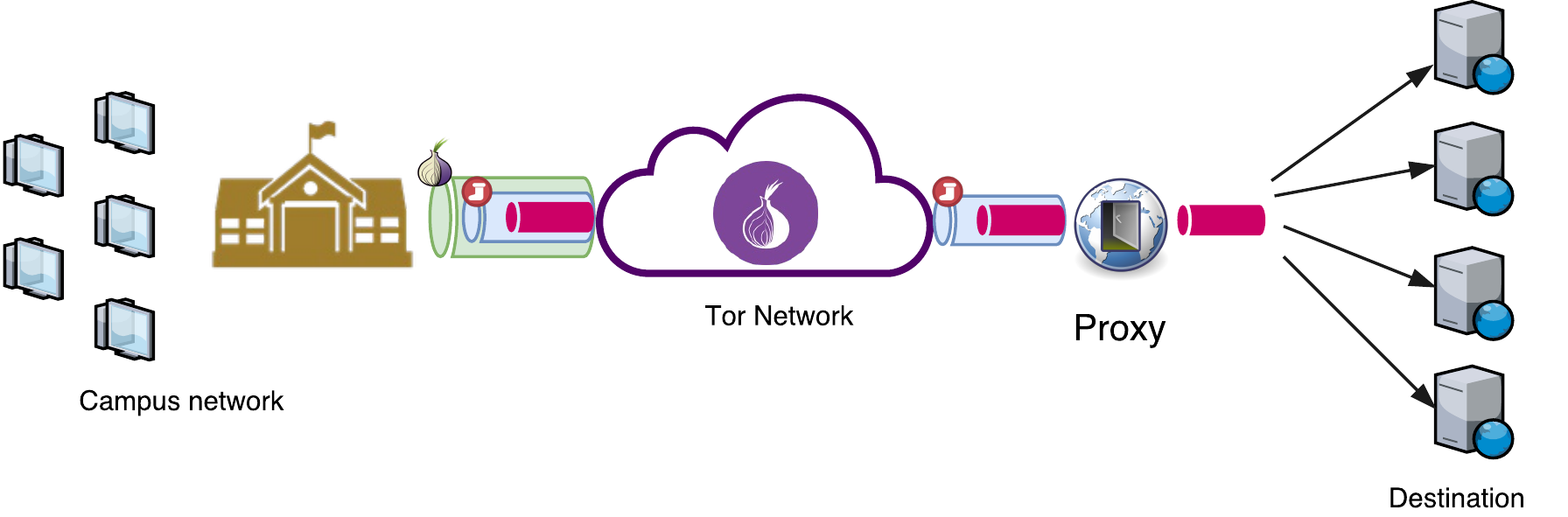}
    \caption{Our experimental setup on Tor}
    \label{fig:exp_setup}
\end{figure*}

Figure~\ref{fig:exp_setup} shows our experimental setup for our Tor experiments. 
We used several Tor clients that we  ran inside separate VMs to generate and collect Tor traffic. 
We use each of our Tor clients to browse the top 50,000 Alexa websites over Tor, and captured the flows entering and exiting the Tor network for these connections (we use half of the connections for training, and the other half for testing). 
Therefore, the entering flows are in Tor cell format, and the flows exiting Tor are in regular HTTP/HTTPS format. 
We used 1,000 arbitrary Tor circuits for browsing  websites over Tor, i.e., each circuit was used to browse roughly 50  websites. 
We used different guard nodes in forming our Tor circuits; we were able to alternate our guard nodes so by disabling Vanilla Tor's option that enforces guard relay reuse. We also used a regular Firefox browser, instead of Tor's browser, to be able to enforce circuit selection. We used Tor version 0.3.0.9, automated by a Python script. 

Note that we did not set up our own Tor relays for the purpose of the experiments, and we merely used public Tor relays in all of our experiments. 
We captured the ingress Tor flows using  \code{tcpdump}  on our Tor clients.  To capture the egress Tor traffic (i.e., traffic from exit relays to websites), 
we made our exit Tor traffic tunnel  through our own SOCKS proxy server (as shown in Figure~\ref{fig:exp_setup}), and we collected the exit Tor traffic on our own SOCKS proxy server using  \code{tcpdump}. Note that using this data collection proxy may add additional latency on the collected flows, so the performance of \name in practice is  better than what we report through experiments. 
We also collected 500 websites through Tor pluggable transport  to evaluate them as countermeasures against \name.

We collected our Tor traffic in two steps: first, we collected traffic over a two weeks period, and then with a three months gap we collected more Tor traffic for a one month period (in order to show the impact of time on training). 
We have made our dataset available publicly. To the best of our knowledge, 
this is largest dataset of correlated Tor flows, and we hope it will be useful to the research community. 

Note that while we only collect web traffic, this is not a constraint of \name, and it can be used to correlate arbitrary Tor traffic.


\subsection{Ethics of Data Collection}

To make sure we did not overload Tor's network, we ran up to 10 concurrent Tor connections during our data collection. 
Also, we alternated the guard nodes used in our circuits  to evade  overloading any specific circuits or relays. 
We did not browse any illegal content over Tor, and 
we used an idle time between  connections of each of our clients. 
As explained above, we collected our ingress and egress Tor flows on our own Tor clients as well as our own SOCKS proxy server; therefore, we did \emph{not} collect any traffic of other Tor users. 

In our experiments with Tor pluggable transports,  we collected a much smaller set of flows compared to our bare Tor experiments; we did so  because Tor bridges are very scarce and expensive, and therefore we avoided overloading the bridges.

\subsection{Choosing the Hyperparameters}


We used  Tensorflow~\cite{abadi2016tensorflow} to implement the neural networks of \name. We tried various values for different hyperparameters of our system to optimize the  flow correlation performance.
To optimize each of the parameters, our network took about a day to converge (we used a single Nvidia TITAN X GPU).

For the learning rate, we tried $\{0.001,0.0001,0.0005,0.00005\}$, and we got the best performance with a learning rate of $0.0001$. 
As for the number of negative samples, $N_{neg}$, we tried $\{9,49,99,199,299\}$ and  $199$ gave us the best results. For the  window sizes of the  convolution layers,  $w_1$ and $w_2$, we tried $\{5,10,20,30\}$. 
Our best results occurred with $w_1=30$ and $w_2=10$. 
We also experimented with $\{2,5,10\}$ for the size of the max pooling, and a max pooling of $5$ gave the best performance. 
Finally, for the number of the kernels, $k_1,k_2$, we tried $\{500, 1000, 2000, 3000\}$, and $k_1=2000$ and $k_2=1000$ resulted in the best performance. 
We present the values of these parameters and other  parameters of the system in Table~\ref{table:net_size}.

\begin{table}
    \caption{\name's hyperparameters optimized to correlate Tor traffic. }
    \centering
    \begin{tabular}{c|c}
        \hline
        Layer & Details \\
        \hline
        \multirow{ 4}{*}{Convolution Layer 1}  & Kernel num: $2000$  \\
        & Kernel size: $(2,30)$   \\
        & Stride: (2,1) \\
        & Activation: Relu \\
        \hline
        \multirow{ 2}{*}{Max Pool 1} & Window size: (1,5)\\
        & Stride: (1,1) \\
        \hline
        \multirow{ 4}{*}{Convolution Layer 2}  & Kernel nume: $1000$   \\
        & Kernel size:  $(2,10)$   \\
        & Stride: (4,1) \\
        & Activation: Relu \\
        \hline
        \multirow{ 2}{*}{Max Pool 2} & Window size: (1,5)\\
        & Stride: (1,1) \\
        \hline
        Fully connected 1 & Size: $3000$, Activation: Relu \\
        \hline
        Fully connected 2 & Size: $800$, Activation: Relu\\
        \hline
        Fully connected 3 & Size: $100$, Activation: Relu\\
        \hline
    \end{tabular}

    \label{table:net_size}
\end{table}

\subsection{Evaluation Metrics}\label{metrics}

Similar to previous studies, we use the \emph{true positive (TP)}  and \emph{false positive (FP)} error rates as the main metrics for evaluating the performance of flow correlation techniques. 
The TP rate measures the fraction of associated flow pairs that are correctly declared to be correlated by \name (i.e., a flow pair ($i$,$j$) where $i$ and $j$ are the segments of the \emph{same} Tor connection, and we have $p_{i,j}>\eta$). 
On the other hand, the FP rate measures the fraction of non-associated  flow pairs that are mistakenly identified as correlated by \name (e.g., when $i$ and $j$ are the segments of two unrelated Tor connections, yet $p_{i,j}>\eta$). 
To evaluate FP,  \name correlates every collected entry flow to every collected exit flow, therefore, we perform about $(25,000-1)^2$ false correlations for each of our experiments (we have $25,000$ Tor connections in  our test dataset).

Note that the detection threshold $\eta$ makes a trade off between the FP and TP rates; therefore we make use of  \emph{ROC curves} to compare \name to other algorithms. 

Finally, in our comparisons with RAPTOR~\cite{raptor}, we additionally use the \emph{accuracy} metric (the sum  of true positive and true negative correlations over all correlations), which is used in the RAPTOR paper. 
To have a fair comparison, we derive the accuracy metric similar to  RAPTOR: 
each flow is declared to be associated  with only \emph{a single} flow out of all evaluated flows, e.g., the flow that results in the maximum correlation metric, $p_{i,j}$.
For the rest of our experiments, each flow can be declared as correlated with arbitrary number of intercepted flows (i.e., any pairs that $p_{i,j}>\eta$), which is a more realistic (and more challenging) setting.

\section{Experiment Results}
In this section we present and discuss our  experimental results.

\subsection{A First Look at the Performance}

\begin{figure}[!t]
    \centering
    \includegraphics[scale=0.34]{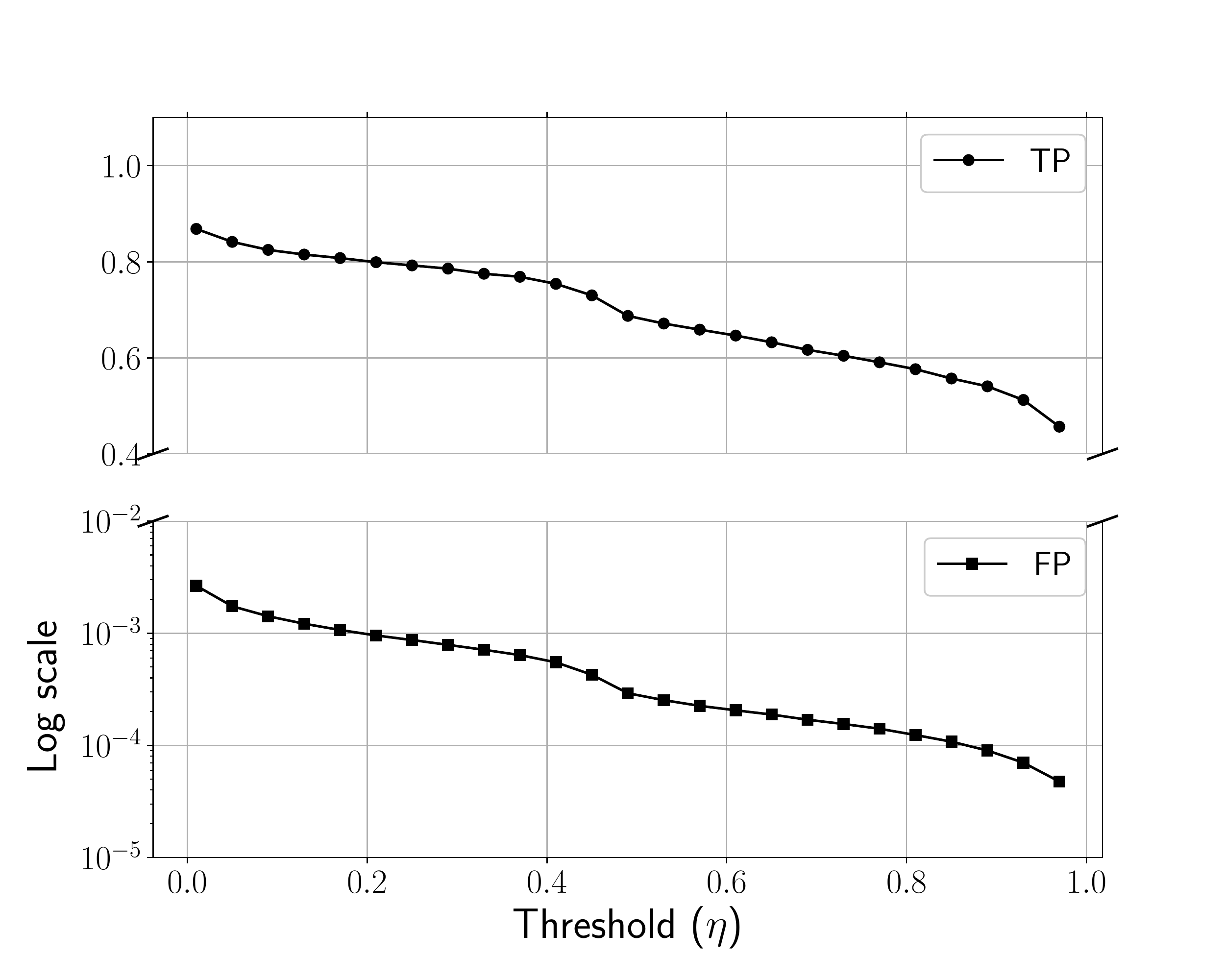}
    \caption{True positive and false positive error rates of \name in detecting correlated pairs of ingress and egress Tor flows for different detection thresholds ($\eta$). Each flow is only 300 packets.}
    \label{fig:300_thresh}
\end{figure}

As described in the experimental setup section, we browse 50,000 top Alexa websites over Tor and collect their ingress and egress flow segments. We use half of the collected traces to train \name (as described earlier). 
Then, we use the other half of the collected flows to test \name. Therefore, we feed \name about $25,000$ pairs of associated flow pairs, and $25,000\times 24,999\approx 6.2\times 10^8$ pairs of non-associated flow pairs for training. 
We only use the first $\ell=300$ packets of each flow (for shorter flows, we pad them to 300 packets by adding zeros). 
Figure~\ref{fig:300_thresh} presents the true positive and false positive error rates of \name for different values of  the threshold $\eta$. As expected, $\eta$ trades off the TP and FP error rates. 
The figure shows a promising performance for \name in correlating Tor flows\textemdash using only  300 packets of each flow. 
For instance, for a FP of $10^{-3}$, \name achieves a TP  close to $0.8$. As shown in the following, this is \emph{drastically better} than the performance of previous work. Note that increasing the length of the flows will increase the accuracy, as shown later.

\subsection{\name Can Correlate Arbitrary Circuits and Destinations}

\begin{figure}[!t]
    \centering
    \includegraphics[scale=0.34]{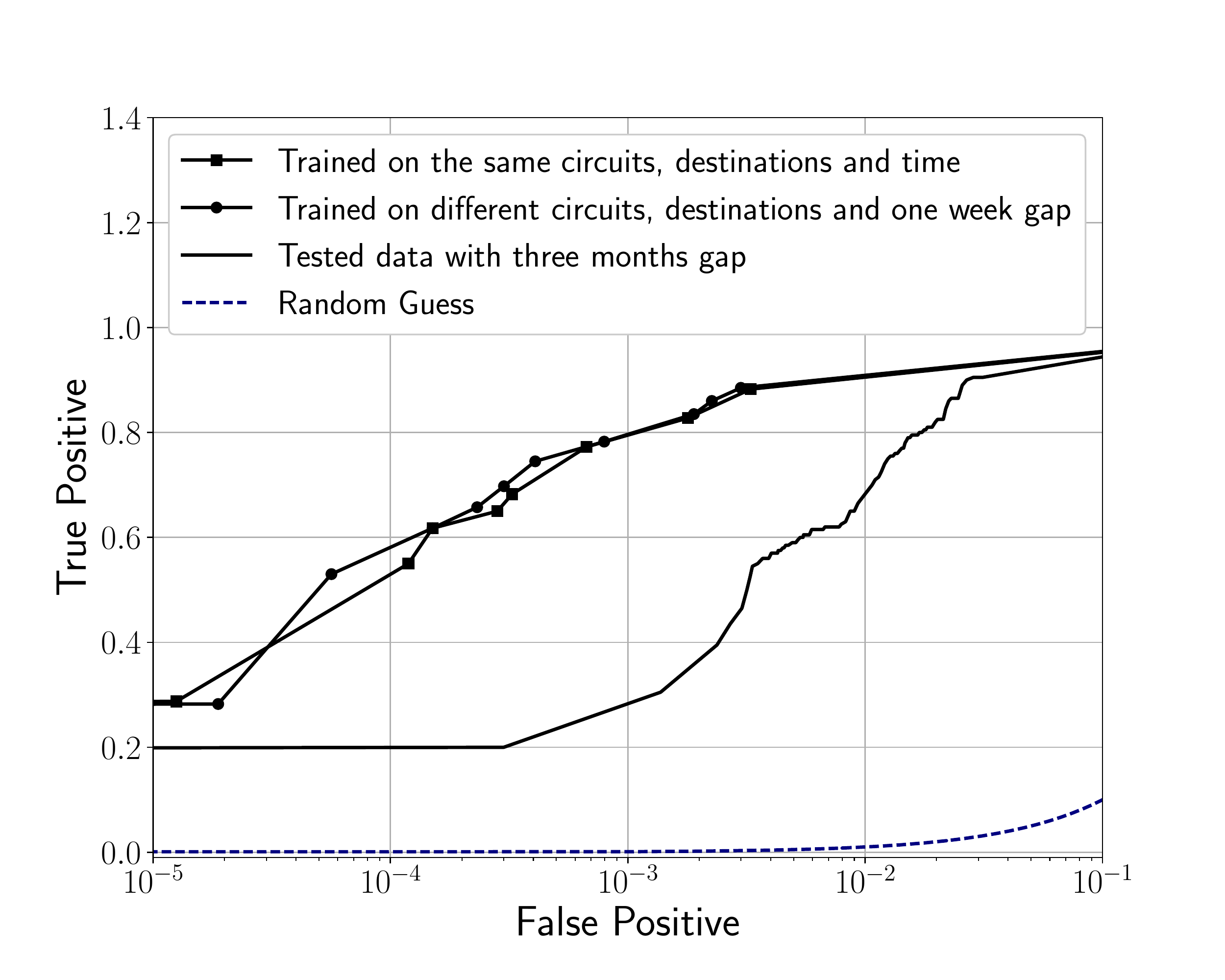}
    \caption{\name's performance does not depend on the circuits and destinations used during the training phase. }
    \label{fig:roc_diff}
\end{figure}

As discussed earlier, \name \emph{learns} a correlation function for Tor that can be used to correlate Tor flows on\textemdash any circuits\textemdash and to\textemdash any destinations\textemdash regardless of the circuits and destinations used during the training process. 
To demonstrate this,  we compare \name's performance in two experiments, each consisting $2,000$ Tor connections, therefore $2,000$ associated pairs and $2,000\times 1,999$ non-associated flow pairs. In the first experiment, the flows tested for correlation by \name use the same circuits and destinations as the flows used during \name's training. In the second experiment, the flows tested for correlation by \name (1) use circuits that are totally different from the circuits used during training, (2) are targeted to web destinations different from those used during training, and (3) are collected one week after  the learning flows.  
Figure~\ref{fig:roc_diff} compares \name's ROC curve for the two experiments. 
As can be seen, \name performs  similarly in both of the experiments, demonstrating that \name's learned correlation function can be used to correlate Tor flows on arbitrary circuits and to arbitrary destinations. 
The third line on the figure shows the results when the training set is three months old, showing a degraded performance, as further discussed in the following.

\subsection{\name Does Not Need to  Re-Train Frequently}\label{sec:exp:traintime}

\begin{figure}[!t]
    \centering
    \includegraphics[scale=0.27]{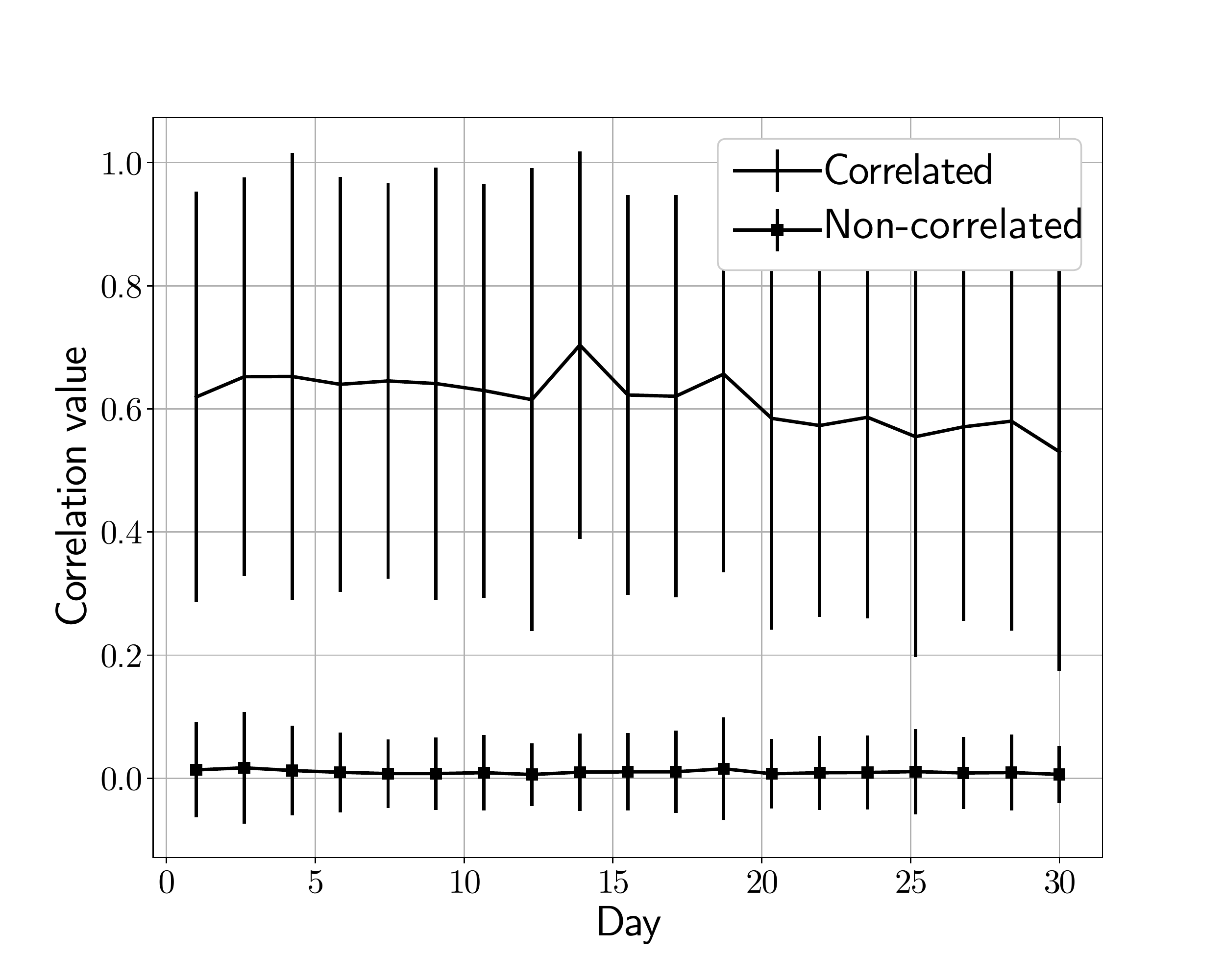}
    \caption{\name's correlation values for associated and non-associated flows for 30 consecutive days without retraining. The performance only starts to drop after about three weeks.}
    \label{fig:onemonth}
\end{figure}

Since the characteristics of Tor traffic change over time, any learning-based algorithm needs to be re-trained occasionally to preserve its correlation performance. 
We performed two experiments to evaluate how frequently \name needs to be retrained. 
In our first experiment,  we evaluated our pre-trained model over Tor flows collected during 30 consecutive  days. Figure~\ref{fig:onemonth} presents the output of the correlation function for each of the days for both associated and non-associated flow pairs. 
 As we can see, the correlation values for non-associated flows do not change substantially, however, the correlation values for associated flows starts to slightly degrade after about three weeks. This suggests that an adversary will need to retrain her \name  \emph{only every three weeks}, or even once a month. 

%

As an extreme case, we also evaluated \name's performance using a model that was trained three months earlier. Figure~\ref{fig:roc_diff} compares the results in three cases: three months gap between training and test, one week gap between training and test, and no gap. 
We see that \name's accuracy significantly degrades with three months gap between training and test\textemdash interestingly, even this significantly degraded performance of \name due to lack of retraining is superior to all previous techniques compared in Figure~\ref{fig:tor_compare}.

\subsection{\name's Performance Does Not Degrade with the Number of Test Flows}

\begin{figure}
    \centering
    \includegraphics[scale=0.34]{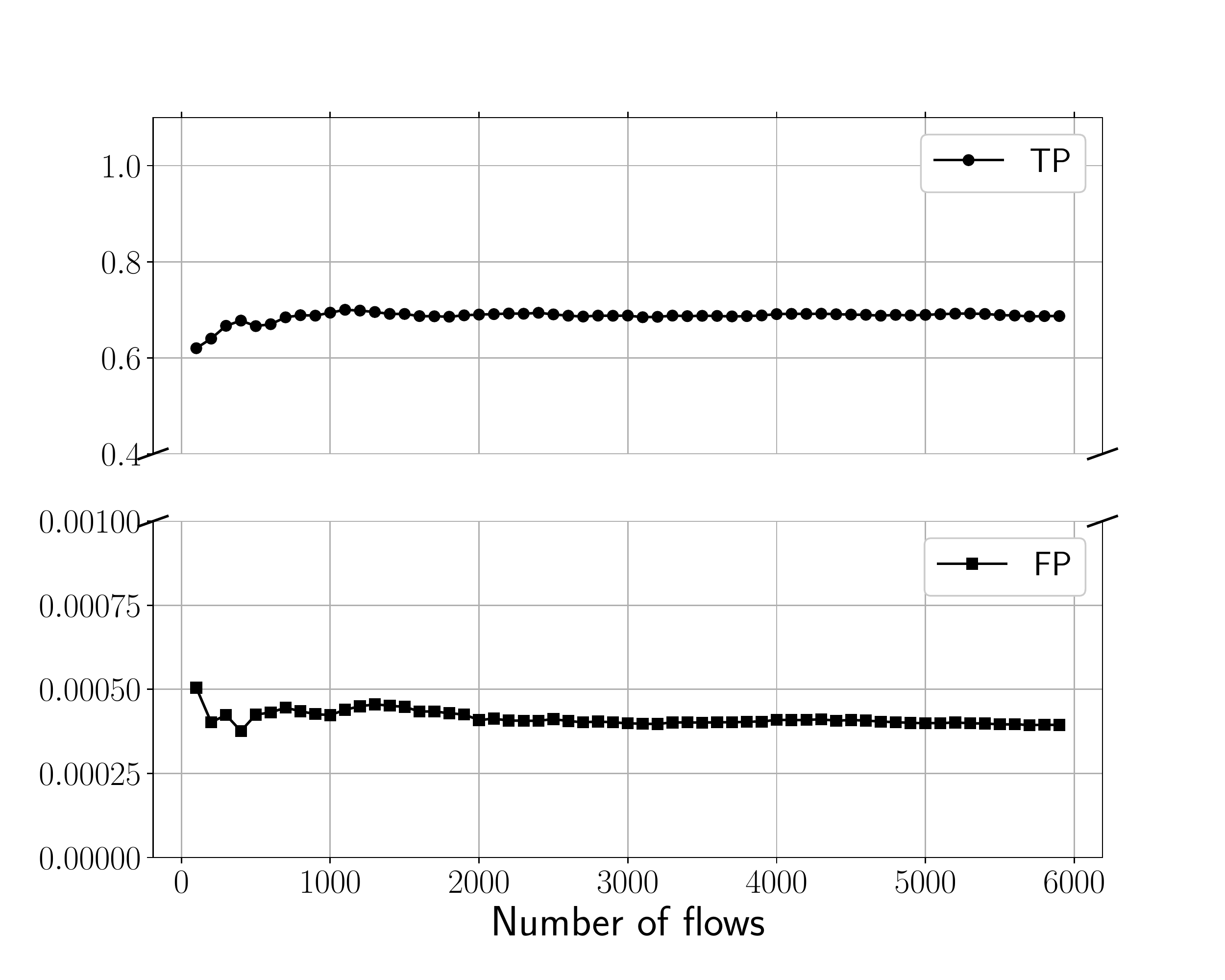}
    \caption{\name's performance is consistent regardless of the size of the testing dataset (we use a fixed, arbitrary $\eta$).}
    \label{fig:diff_sizes}
\end{figure}

We also show that \name's correlation performance does not depend on the number of flows being correlated, i.e., the size of the test dataset.   
Figure~\ref{fig:diff_sizes} presents the TP and FP results (for a specific threshold) on datasets with different numbers of flows. As can be seen, the results are consistent for different numbers of  flows being correlated.
This suggests that \name's  correlation performance  will be similar to what derived through our experiments  even if \name is applied on significantly larger datasets of  intercepted flows, e.g., on the flows collected by a large malicious IXP.    

\subsection{\name's Performance Rapidly Improves with Flow Length}\label{sec:res:performance:flow-length}

\begin{figure}[!t]
    \centering
    \includegraphics[scale=0.34]{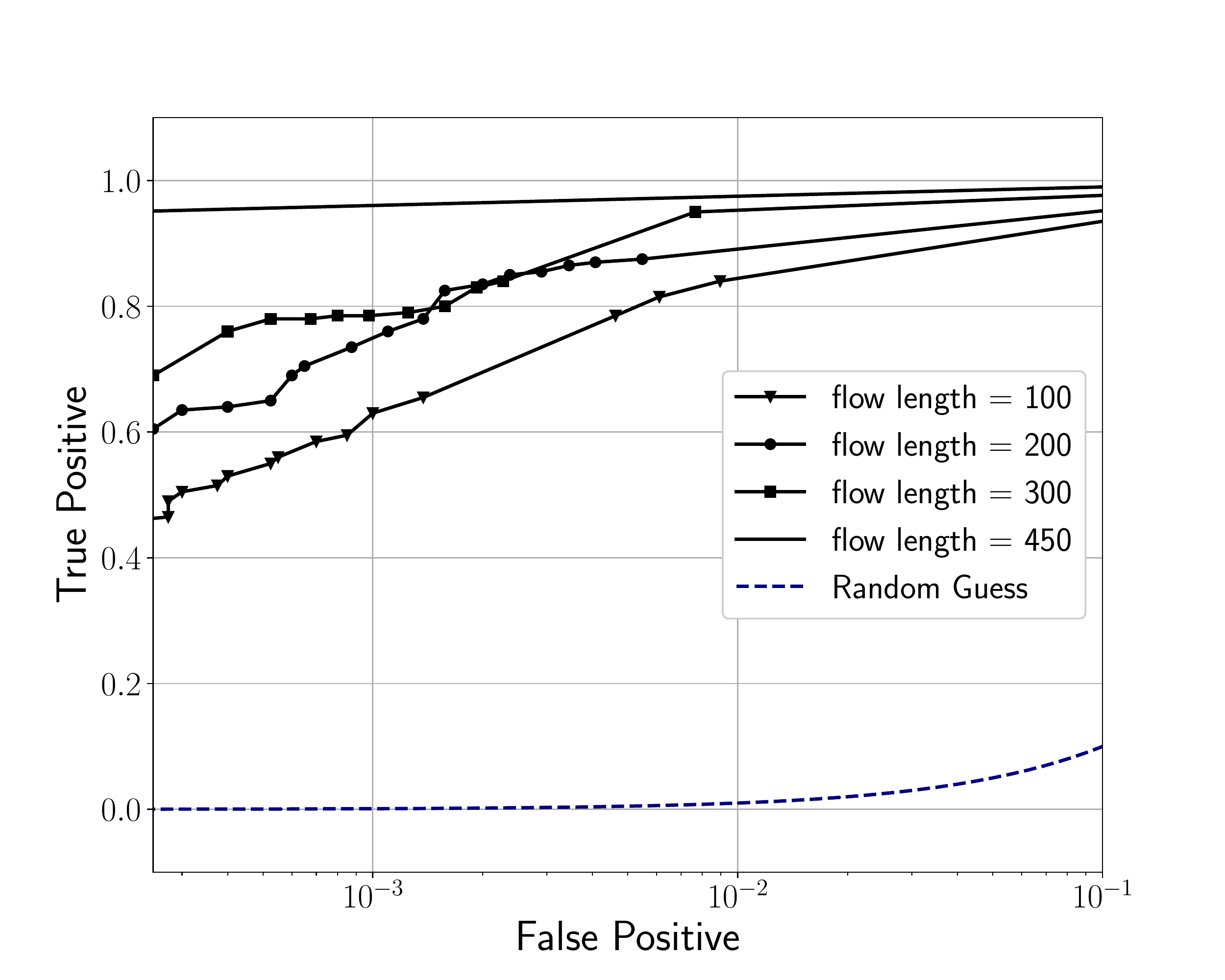}
    \caption{\name's performance rapidly improves when using longer flows for training and testing. }
    \label{fig:diff_flowsize}
\end{figure}

In all of the previous results, we used a flow length of $\ell=300$ packets. 
As can be expected, increasing the length of the flows used for training and testing should improve the performance of \name. 
Figure~\ref{fig:diff_flowsize} compares  \name's performance for different lengths of flows, showing that \name's performance improves significantly for longer flow observations.  For instance, for a target FP of $10^{-3}$, \name achieves $TP=0.62$ with $\ell=100$ packets long flows, while it achieves  $TP=0.95$ with flows that contain $\ell=450$ packets.  

Note that the lengths of intercepted flows makes a tradeoff between \name's performance and the adversary's computation overhead.  
That is, while a larger flow length improves \name's correlation performance, longer flows impose higher  storage and computation overheads on the traffic correlation adversary. A larger flow length  also increase the adversary's waiting time in detecting correlated  flows  in real-time.

\subsection{\name's Performance Improves with the Size of the Training  Set}

\begin{figure}
    \centering
    \includegraphics[scale=0.34]{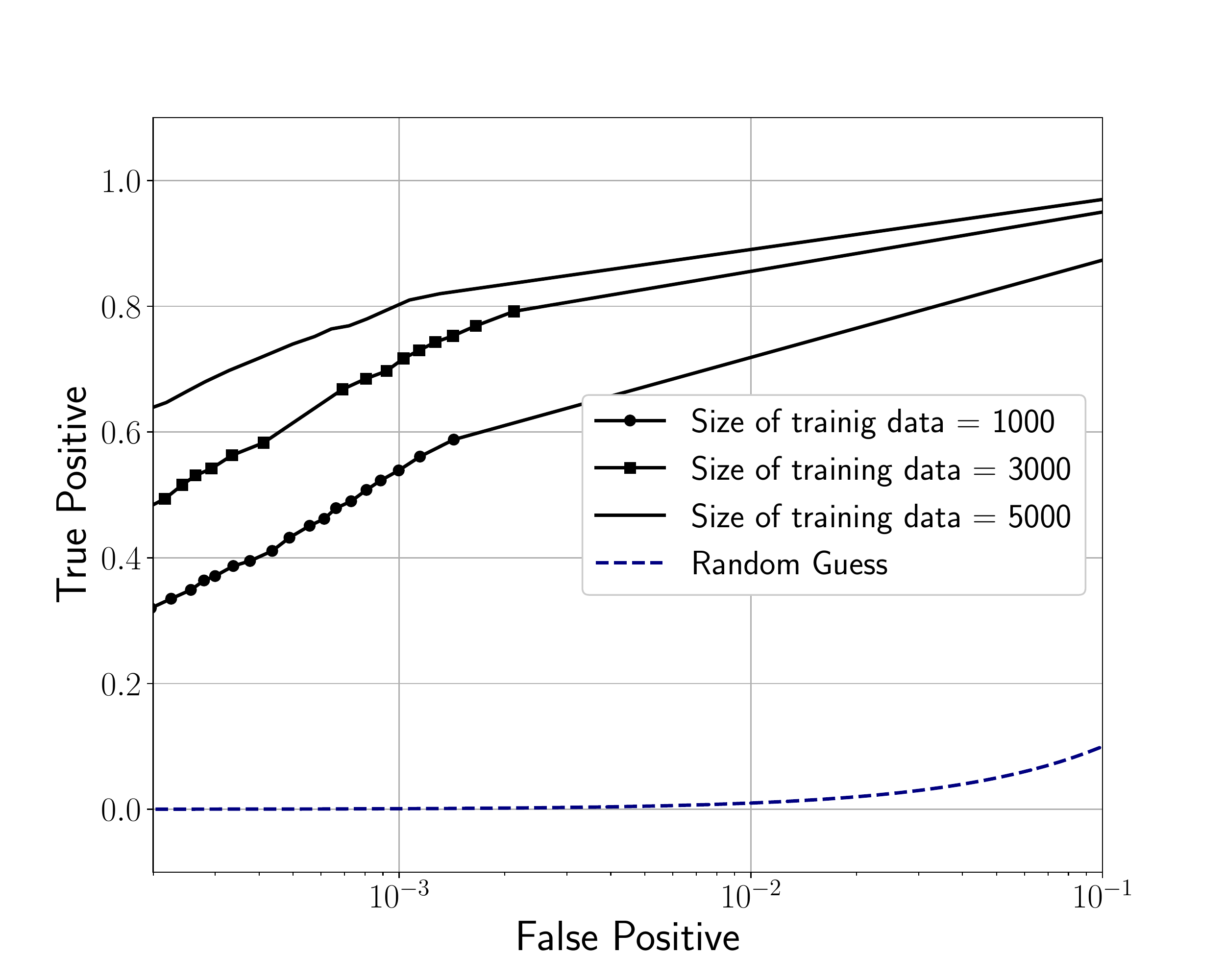}
    \caption{\name's correlation performance improves with more training data. }
    \label{fig:diff_train-data}
\end{figure}

As intuitively expected, \name's performance improves when it uses a larger set of Tor flows during the training phase (i.e., \name learns a better correlation function for Tor with more training samples).
   Figure~\ref{fig:diff_train-data} compares \name's ROC curve when trained with different numbers of flows (for all of the experiments, we use a fixed number of 1,000 flows for testing). The figure confirms that increasing the size of the training set improves the performance of \name. For instance, for a target $FP=10^{-3}$, using 1,000 training flows results in $TP=0.56$, while using 5,000 flows for training gives \name a $TP=0.8$.
This shows that a resourceful adversary can improve the accuracy of her flow correlation classifier by collecting a larger number of Tor flows for training. Note that a larger training set increases the training time, however  the learning process does not need to repeat frequently as evaluated before.

\subsection{\name Significantly Outperforms the State-Of-The-Art}

\begin{figure}
    \centering
    \includegraphics[scale=0.32]{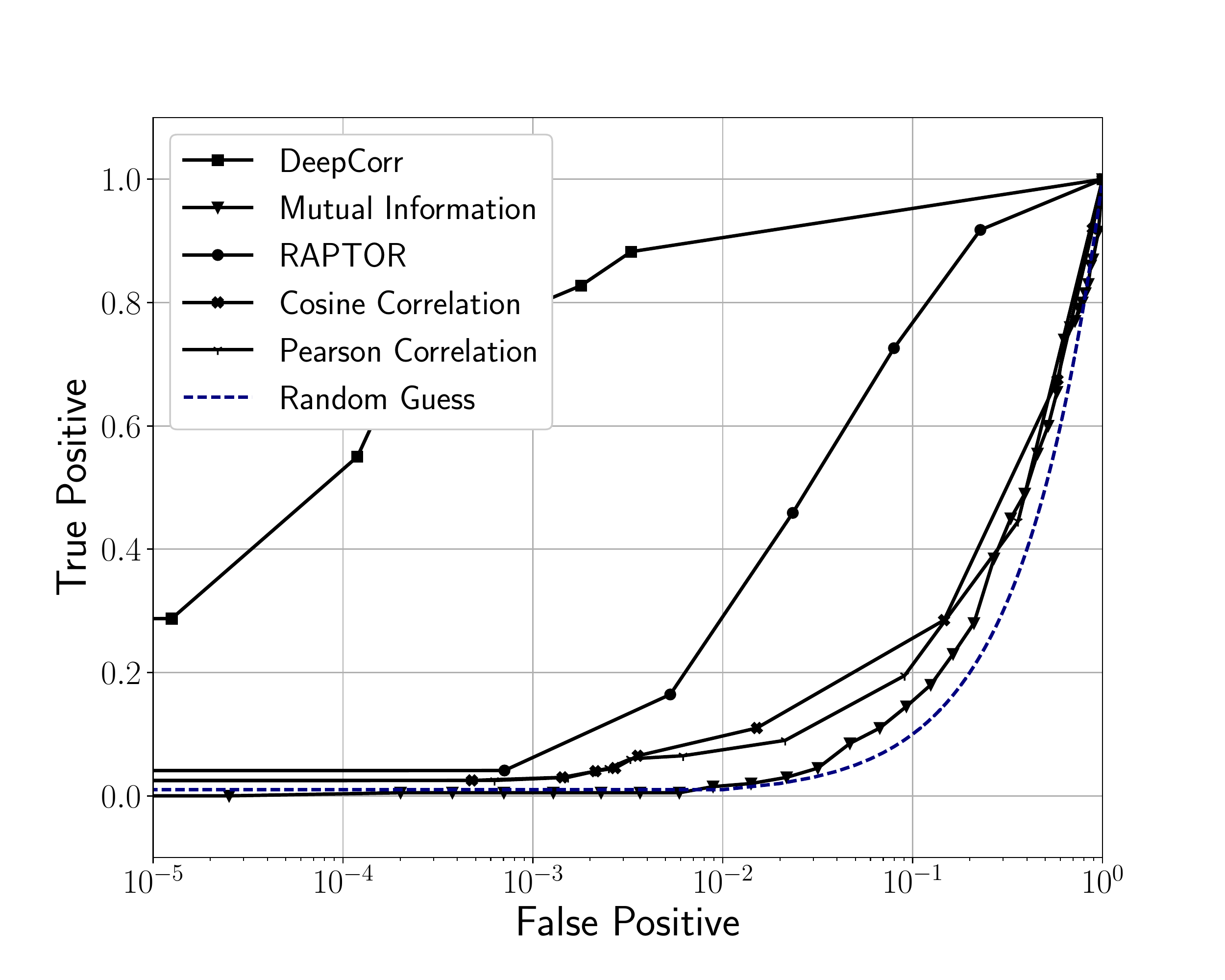}
    \caption{Comparing \name's ROC curve with previous systems shows an overwhelming improvement over the state-of-the-art (all the systems are tested on the same dataset of flows, and each flow is 300 packets).}
    \label{fig:tor_compare}
\end{figure}

In Section~\ref{sec:prev-tech} we overviewed  major flow correlation techniques introduced prior to our work. We perform experiments to compare \name's performance with such prior systems in correlating Tor flows.
Figure~\ref{fig:tor_compare} compares the ROC curve of \name to other systems, in which all the systems are tested on the exact same set of Tor flows (each flow is at most 300 packets).    
As can be seen, \emph{\name significantly outperforms the flow correlation algorithms used by prior work}, as we see a wide gap between the ROC curve of \name and other systems. 
For instance, for a target $FP=10^{-3}$, while \name achieves a TP of $0.8$, previous systems provide TP rates less than $0.05$! This huge improvement comes from the fact that \name learns a correlation function tailored to Tor whereas previous systems use generic statistical correlation metrics (as introduced in Section~\ref{sec:prev-tech}) to link Tor connections. 

\begin{figure}
    \centering
    \includegraphics[scale=0.3]{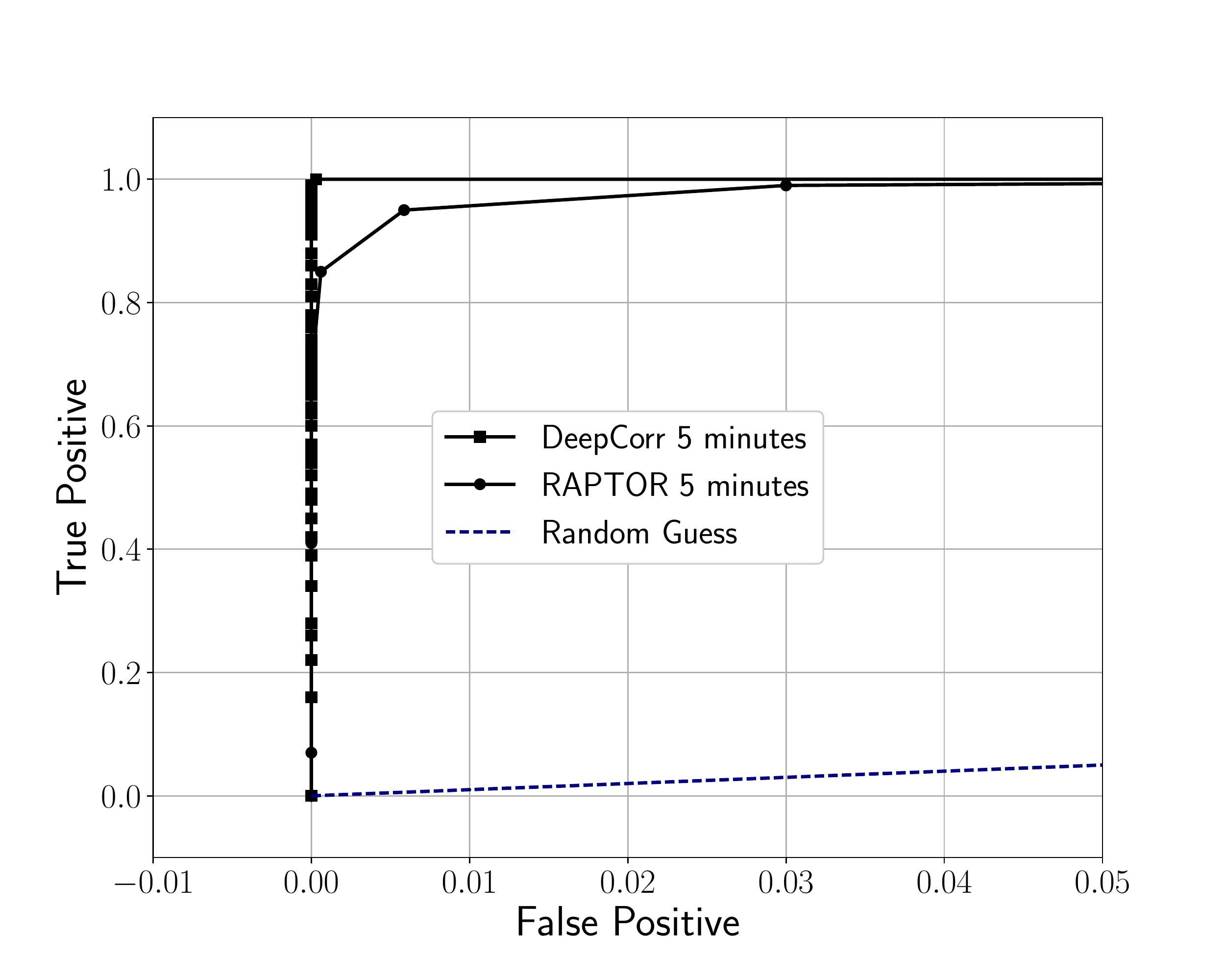}
    \caption{Comparing \name to RAPTOR~\cite{raptor} using the same flow lengths and flow number as the RAPTOR~\cite{raptor} paper.  }
    \label{fig:tor_compare_raptor}
    \end{figure}

\begin{figure}
\centering
    \includegraphics[scale=0.3]{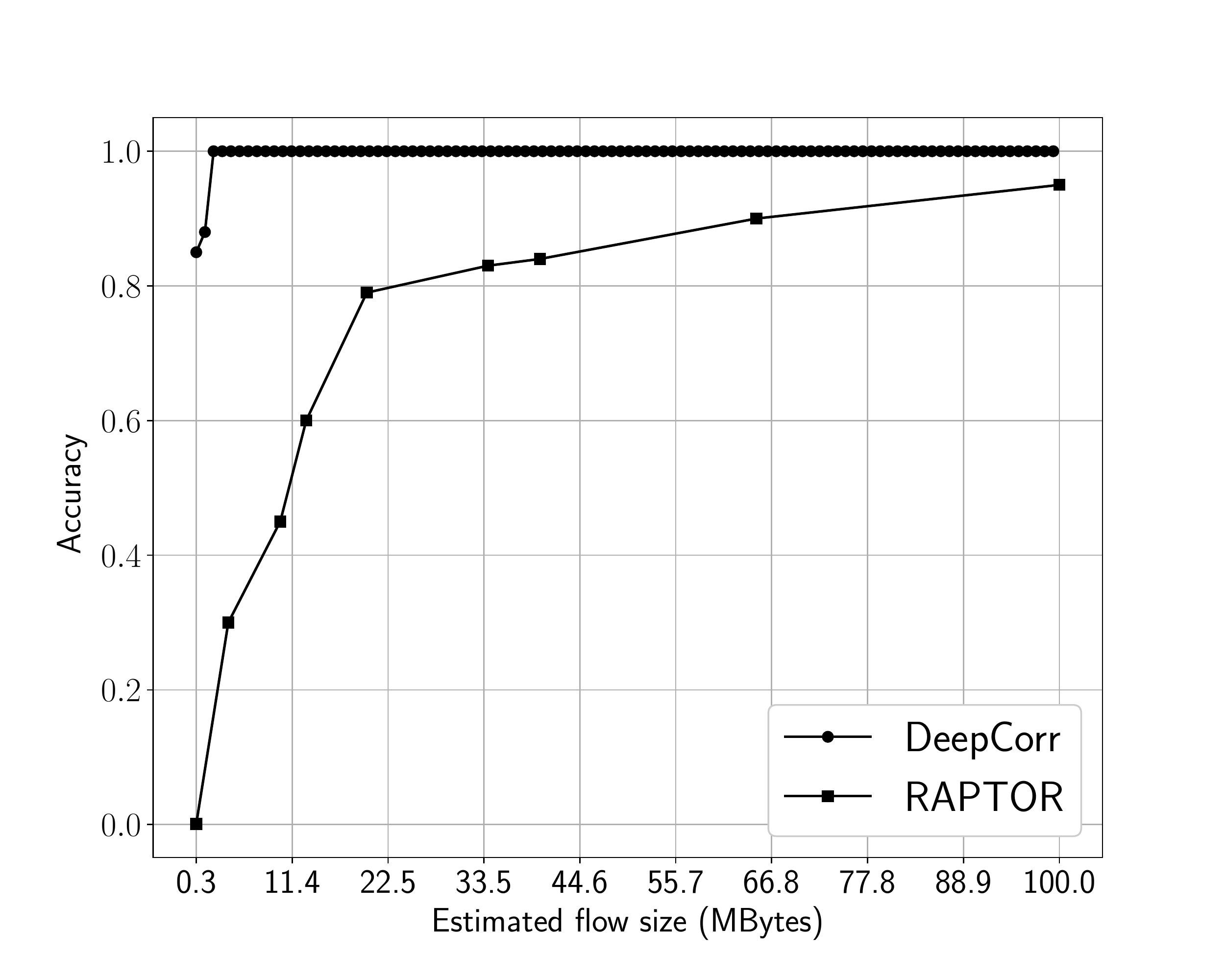}
    \caption{Comparing the accuracy of \name and RAPTOR~\cite{raptor} for various volumes of data intercepted from each flow. The RAPTOR values are comparable to Figure~6 of the RAPTOR paper~\cite{raptor}. }
    \label{fig:acc_compare2}

\end{figure}

\begin{figure}
    \centering
    \includegraphics[scale=0.3]{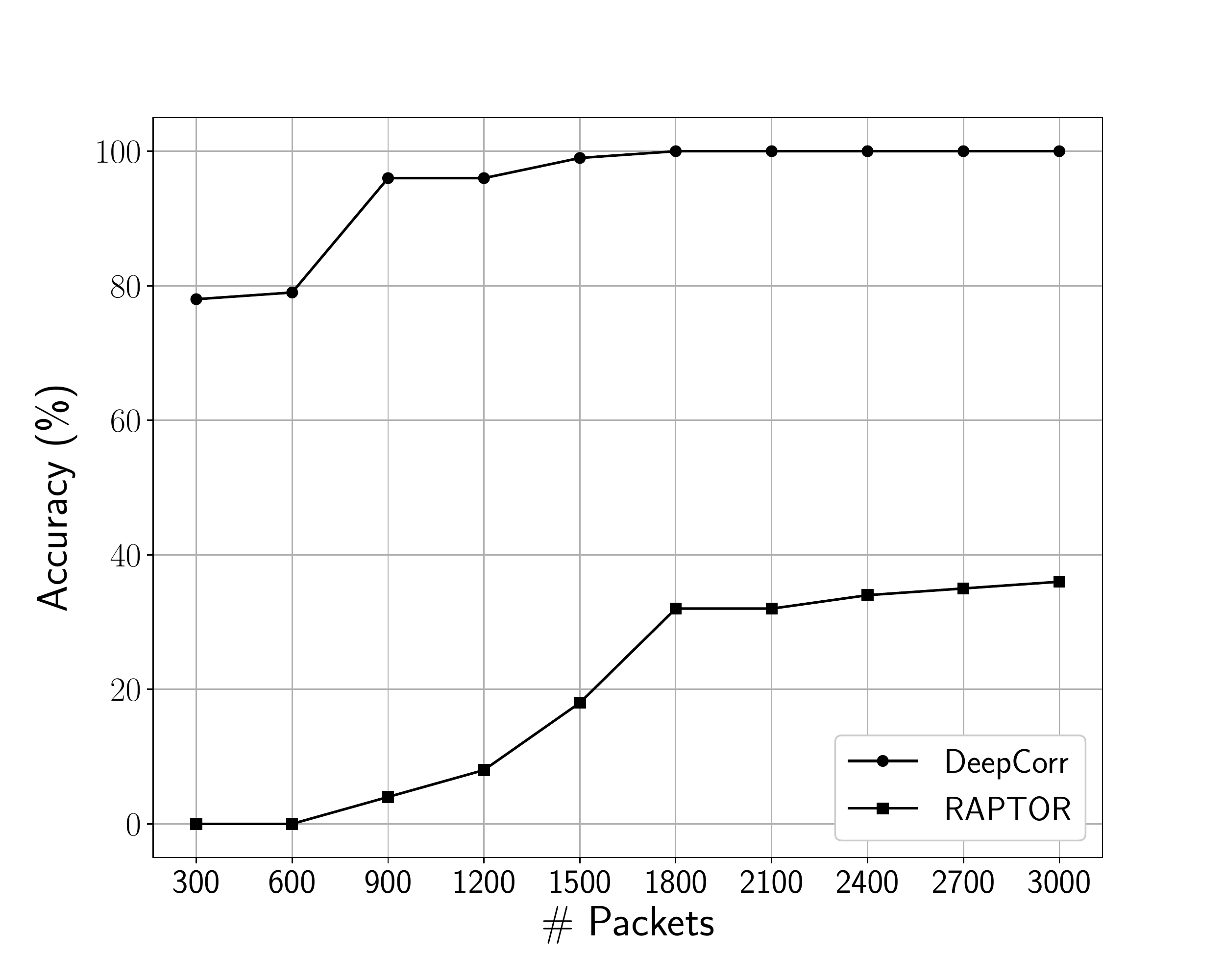}
    \caption{Comparing \name to RAPTOR in correlating  short flows. }
    \label{fig:acc_compare}
\end{figure}

Needless to say, any flow correlation algorithm will improve its performance by increasing the length of the flows it intercepts for correlation (equivalently, the  traffic volume it collects from each flow); we showed this in   Section~\ref{sec:res:performance:flow-length} for \name. 
To offer reasonable accuracies, previous works have performed their experiments on flows that contain  significantly more packets (and more data) than our experiments. 
For instance, Sun et al.\ evaluated the state-of-the-art RAPTOR~\cite{raptor} in a setting with \emph{only 50 flows}, and each flow carries \emph{100MB of data over 5 minutes}. 
This is while in our  experiments presented so far,  each flow has only 300 packets, which is equivalent to only $\approx 300$ KB of Tor traffic (in contrast to RAPTOR's 100MB!).
To ensure a fair comparison, we evaluate \name to RAPTOR in the exact same setup (e.g., 50 flows each 100MB, and we use the accuracy metric  described in Section~\ref{metrics}). The results shown in Figure~\ref{fig:tor_compare_raptor} demonstrates  \name's drastically superior performance (our results for RAPTOR comply with the numbers reported by   Sun et al.~\cite{raptor}).
On the other hand, we show that the performance gap between \name and RAPTOR is significantly wider for shorter flow observations. 
To show this, we  compare \name and RAPTOR based on the volume of traffic they intercept from each flow. The results shown  in 
Figure~\ref{fig:acc_compare2} demonstrate that \name outperforms significantly, especially for shorter flow observations. 
For instance, RAPTOR achieves a $0.95$ accuracy after receiving 100MB from each flow, whereas \name achieves an accuracy of $1$  with about 3MB of traffic. 
We see that \emph{\name is particularly powerful on shorter flow observations}. 
We zoomed in by comparing RAPTOR and \name for small number of observed packets, which is shown in Figure~\ref{fig:acc_compare}. We see that \name achieves an accuracy of $\approx 0.96$ with only 900 packets, in contrast to RAPTOR's 0.04 accuracy.


\subsection{\name's Computational Complexity}

\begin{table}
    \caption{Correlation time comparison with previous techniques}
    \label{table:time_compare}
    \begin{tabular}{c|c}
        Method & One correlation time \\
        \hline
        RAPTOR & $0.8ms$ \\
        Cosine  & $0.4ms$ \\
        Mutual Information & $1ms$ \\
        Pearson& $0.4ms$ \\       
        DeepCorr & $2ms$ \\
        \hline
    \end{tabular}
\end{table}

In Table~\ref{table:time_compare}, we show the time to perform a single \name correlation in comparison to  that of previous techniques (the  correlated flows are 300 packets long for all the systems). 
We see that \name is noticeably slower that previous techniques, e.g., roughly two times slower than RAPTOR. 
However, note that since all the systems use the same length of flows, \emph{\name offers drastically better correlation performance for the same time overhead}; for instance, based on  Figure~\ref{fig:tor_compare}, we see that \name offers
a TP$\approx 0.9$ when all previous systems offer a TP less than $0.2$.
Therefore, when all the systems offer similar accuracies (e.g.,  each using various lengths of input flows) \name will be \emph{faster} than all the systems for the same accuracy.  As an example, each RAPTOR correlation takes 20ms (on much longer flow observations) in order to achieve the same accuracy as \name which takes only 2ms\textemdash i.e., \name is 10 times faster for the same accuracy.

Compared to previous correlation techniques, \name is the only system that has a training phase. 
We trained \name  using a standard Nvidia TITAN X GPU (with 1.5GHz clock speed and 12GB of memory) on about 25,000 pairs of associated flow pairs and $25,000\times 24,999\approx 6.2\times 10^8$ non-associated flow pairs, where each flow consists of 300 packets. 
In this setting, \name is trained in roughly one day. 
Recall that as demonstrated in Section~\ref{sec:exp:traintime}, \name does not need to be re-trained frequently, e.g., only once every three weeks. 
Also, a resourceful adversary with better  GPU resources than ours will be able to cut down on the training time. 



\subsection{\name Works in Non-Tor Applications as Well}

While we presented \name as a flow correlation attack on Tor, it can be used to correlate flows in other flow correlation applications as well. 
We demonstrate this by applying \name to the problem of  stepping stone attacks~\cite{he2007detecting,wang2003robust,blum2004detection}. 
In this setting, a cybercriminal proxies her traffic through a compromised machine (e.g., the stepping stone) in order to hide her identity. Therefore, a network administrator can use flow correlation to match up the ingress and egress segments of the relayed connections, and therefore trace back to the cybercriminal. Previous work has devised various flow correlation techniques for this application~\cite{nasr2017compressive,houmansadr:ndss09,paxson:ton95,wang:esorics02,donoho:raid02}. 

\begin{figure}
    \centering
    \includegraphics[scale=0.12]{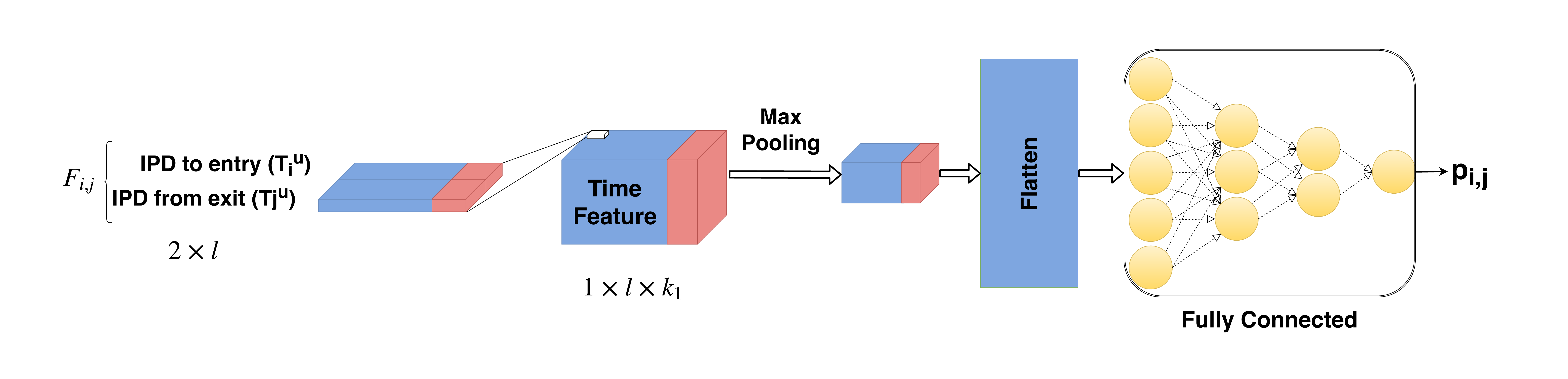}
    \caption{The network architecture of \name to detect stepping stone attacks}
    \label{fig:stepping_model}
\end{figure}

For our stepping stone detection experiments, we used the 2016 CAIDA anonymized data traces~\cite{caida_trace_2016}. Similar to the previous works~\cite{nasr2017compressive,houmansadr:ndss09,houmansadr2014non}  we simulated the network jitter using Laplace distribution, and modeled packet  drops by a Bernoulli distribution with different rates.
We apply \name to this problem by learning \name in a stepping stone setting.  
As the noise model is much simpler in this scenario than Tor, we use a simpler neural network model for \name for this application. Also, we only use one direction of a bidirectioal connection to have  a fair comparison with previous systems, which all only use  one-sided flows.  
Figure~\ref{fig:stepping_model} and Table~\ref{table:net_size_stepping} show our tailored neural network and our choices of parameters, respectively.

\begin{table}
    \caption{\name's parameters optimized for the stepping stone attack application. }
    \centering
    \begin{tabular}{c|c}
        \hline
        Layer & Details \\
        \hline
        \multirow{ 4}{*}{Convolution Layer 1}  & Kernel num: $200$  \\
        & Kernel size: $(2,10)$   \\
        & Stride: (1,1) \\
        & Activation: Relu \\
        \hline
        \multirow{ 2}{*}{Max Pool 1} & Window size: (1,5)\\
        & Stride: (1,1) \\
        \hline
        Fully connected 1 & Size: $500$, Activation: Relu \\
        \hline
        Fully connected 2 & Size: $100$, Activation: Relu\\
        \hline
    \end{tabular}

    \label{table:net_size_stepping}
\end{table}

\begin{figure}[!t]  
    \centering
\includegraphics[scale=0.3]{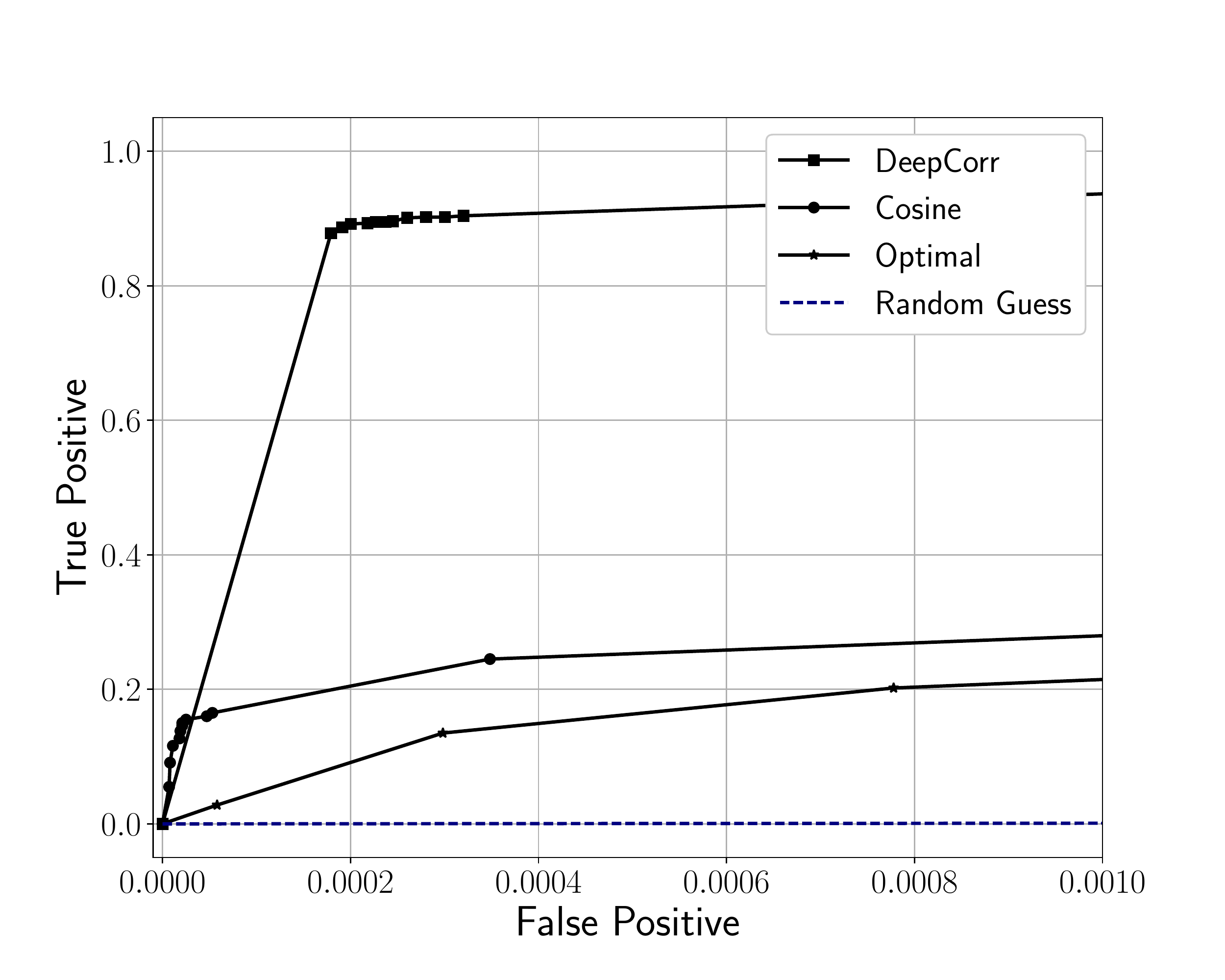}
\caption{\name outperforms state-of-the-art stepping stone detectors in noisy networks (1\% packet drop rate).
}
\label{fig:compare_opt_drop}
\end{figure}



Our evaluations show that 
 \name provides a performance comparable to ``Optimal'' flow correlation techniques of Houmansadr et al.~\cite{houmansadr:ndss09,houmansadr2014non} when network conditions are stable. 
However, when the network conditions becomes noisy, \name offers a significantly stronger performance in detecting stepping stone attacks. This is shown in Figure~\ref{fig:compare_opt_drop}, where the communication network has a network jitter with a 0.005s  standard deviation, and  the network randomly drops 1\% of the packets. 


\section{Countermeasures}

While previous work has studied different countermeasures against flow correlation~\cite{pluggable-transport,obfsproxy,skypemorph,freewave}, they remain mostly non-deployed presumably due to the poor performance of previous flow correlation techniques at large scale~\cite{critic-fing,PETs17-TagIt}. 
In the following we discuss two possible countermeasures.

\subsection{Obfuscate Traffic Patterns}

An intuitive countermeasure against flow correlation (and similar traffic analysis attacks like website fingerprinting) is to \emph{obfuscate traffic characteristics} that are used by such algorithms. Therefore, various countermeasures have been suggested that modify packet timings and packet sizes to defeat flow correlation, in particular by padding or splitting packets in order to modify packet sizes, or by delaying packets in order to perturb their timing characteristics. 
The Tor project, in particular, has deployed various pluggable transports~\cite{pluggable-transport} in order to defeat censorship by  nation-states who block all  Tor traffic. 
Some  of these pluggable transports only obfuscate packet contents~\cite{obfsproxy}, some of them obfuscate the IP address of the Tor relays~\cite{meek}, and some obfuscate traffic patterns~\cite{obfsproxy,skypemorph}. 
Note that Tor's pluggable transports are designed merely for the purpose of censorship resistance, and they obfuscate traffic only from a censored client to her first Tor relay (i.e., a Tor bridge). Therefore, Tor's pluggable transports are \emph{not} deployed by any of Tor's public relays. 

As a possible countermeasure against \name, we suggest to  deploy traffic obfuscation techniques by \emph{all} Tor relays (including the guard and middle relays). 
We evaluated the impact of several Tor pluggable transports on \name's performance. 
Currently, the Tor project has three deployed plugs: meek, obfs3, and obs4. We evaluated \name on meek and obfs4 (obfs3 is an older version of obfs4). We also evaluated two modes of obfs4: one with IAT mode ``on''~\cite{obfs4-obf}, which obfuscates traffic features, and one with the IAT mode ``off'', which does not obfuscate traffic features. 
We used \name to learn and correlate traffic on these plugs. However, \emph{due to ethical reasons, we collected a much smaller set of flows} for these experiments compared to our previous experiments; this is because Tor bridges are very scarce and expensive, and we therefore avoided overloading the bridges.\footnote{Alternatively, we could set up our own Tor bridges for the experiments. We decided to use real-world bridges to incorporate the impact of actual traffic loads in our experiments. } 
Consequently, our correlation results are very optimistic due to their small training datasets (e.g., a real-world adversary will achieve much higher correlation accuracies with adequate training). 
We browsed 500 websites over obfs4 with and without the IAT mode on, as well as over  meek. 
We trained \name on only 400 flows (300 packets each) for each transport  (in contrast to 25,000 flows in our previous experiments), and tested on another 100 flows. 
Table~\ref{table:transport} summarizes the results. 
We see that meek and obfs4 with IAT=0 provide no protection to \name; note that a 0.5 TP is comparable to what we get for bare Tor if trained on only 400 flows (see Figure~\ref{fig:diff_train-data}), therefore we expect correlation results similar to bare Tor with a larger training set. 
The results are intuitive: meek merely obfuscates a bridge's IP and does not deploy traffic obfuscation (except for adding natural network noise). Also obfs4 with IAT=0 solely obfuscates packet contents, but not traffic features. 
On the other hand, we see that \name has a significantly lower performance in the presence of obfs4 with IAT=1 (again, \name's accuracy will be higher for a real-world adversary who collects more training flows).

Our results suggest that (public) Tor relays should deploy a traffic obfuscation mechanism like obfs4 with IAT=1 to resist advanced flow correlation techniques like \name. However, this is \emph{not} a trivial solution due to the increased cost, increased overhead (bandwidth and CPU), and reduced QoS imposed by such obfuscation mechanisms. Even the majority~\cite{obfs4-obf} of Obfsproxy Tor bridges run obfs4 without traffic obfuscation (IAT=0). 
Therefore, designing an obfuscation mechanism tailored to Tor that makes the right balance between performance, cost, and anonymity remains  a challenging  problem for future work.

%
  
  \begin{table}\centering
        \caption{\name's performance if Tor's pluggable transports are deployed by the relays (results are very optimistic due to our small training set, which is for ethical reasons).}
      \begin{tabular}{p{3cm}|c|c}
        Plug name   & TP & FP \\
        \hline
        obfs4 with IAT=0 & $\approx 0.50$ & $0.0005$ \\          
        \hline
        meek  & $\approx$ 0.45 & $0.0005$ \\
        \hline
        obfs4 with IAT=1 & $\approx .10$ & $0.001$ \\          
      \end{tabular}
      \label{table:transport}
  \end{table}


\subsection{Reduce An Adversary's Chances of Performing  Flow Correlation}

Another countermeasure against flow correlation on Tor is reducing an adversary's chances of 
intercepting the two ends of many Tor connections (therefore, reducing her chances of performing flow correlation).
As discussed earlier, recent studies~\cite{feamster:wpes04,raptor,murdoch:pet07} show that various ASes and IXPs intercept a significant fraction of Tor traffic, putting them in an ideal position to perform flow correlation attacks. 
To counter, several proposals suggest  new relay selection mechanisms for Tor that reduce the interception chances of malicious ASes~\cite{akhoondi2012lastor,juen2015defending,nithyanand2015measuring,barton2016denasa,sun2017counter,tan2016data}. None of such alternatives have been deployed by Tor due to their negative  impacts on  performance, costs, and privacy. 
We argue that designing  practical AS-aware relay selection mechanisms for Tor is a promising avenue to defend against flow correlation attacks on Tor.

\section{Conclusions}

We design a flow correlation  system, called \name, that drastically outperforms the state-of-the-art  systems  in correlating Tor connections. \name leverages an advanced  deep learning architecture to \emph{learn} a flow correlation function tailored to Tor's complex network (as opposed to previous works' use of general-purpose statistical correlation metrics). We show that with adequate learning, \name can correlate Tor connections (and therefore break its anonymity) with accuracies significantly stronger than existing algorithms, and using substantially shorter lengths of flow observations.   
We hope that our work demonstrates the escalating  threat of flow correlation attacks on Tor in rise of advanced learning algorithms, and calls for the deployment of effective  countermeasures by the Tor community.  

\section*{Acknowledgments}
This work was supported  by the  NSF grants CNS-$1525642$, CNS-$1553301$, and CNS-$1564067$.

\bibliographystyle{acm}
\bibliography{./bibs}


\end{document}